\documentclass[review,3p,sort&compress]{elsarticle}
\usepackage{amsmath,bm}
\usepackage{adjustbox}
\usepackage{graphicx}
\usepackage{lineno,hyperref}
\modulolinenumbers[5]

\journal{Journal of \LaTeX\ Templates}

\begin{document}

\begin{frontmatter}

\title{Effect of Inner Crust EoS on Neutron star properties.\tnoteref{Inner Crust}}


\author[1]{Ishfaq A. Rather\corref{cor1}}
\ead{ishfaqrather81@gmail.com}
\author[1]{A. A. Usmani}
\ead{anisul.usmani@gmail.com}
\author[2,3]{S. K. Patra}
\ead{patra@iopb.res.in}
\cortext[cor1]{Corresponding author}

\address[1]{Department of Physics, Aligarh Muslim University, Aligarh 202002, India}
\address[2]{Institute of Physics, Bhubaneswar 751005, India}
\address[3]{ Homi Bhabha National Institute, Training School Complex, Anushakti Nagar, Mumbai 400094, India}



\begin{abstract}
	The neutron star maximum mass and the radius are investigated within the framework of the relativistic mean-field (RMF) model. The variation in the radius at the canonical mass, $R_{1.4}$, using different inner crust equation of state (EoS) with different symmetry energy slope parameter is studied. It is found that although the NS maximum mass and the corresponding radius do not vary much with different inner crust EoSs, the radius and the tidal deformability at 1.4$M_{\odot}$ vary with the different choice of crust EoS and variation of about 1-2 km is seen in the radius at the canonical mass. For non-unified EoSs, the crust with a low symmetry energy slope parameter produces a low NS radius at the canonical mass. The properties of maximally rotating neutron stars are also studied. The variation in the radius of rotating star at the canonical mass 1.4$M_{\odot}$ is also seen with the slope parameter.  Similar to the static neutron star, the radius at 1.4$M_{\odot}$ of rotating neutron star is affected by slope parameter of the inner crust. Other important quantities like moment of inertia, frequency, rotational kinetic energy to gravitational energy ratio are also calculated. The variation in these quantities with the crust slope parameter is found to be more proportional to the mass and the radius of NS.

\end{abstract}


\end{frontmatter}


\section{Introduction}
\label{intro}
The structure and the properties of Neutron Stars (NS) have been studied effectively from experimental as well as theoretical models. Such studies reveal the inner structure of NS and the presence of exotic phases. The results obtained from the astrophysical observations require several theoretical inputs for the interpretation. A coextensive effort from theory and experiments have improved and provided new insights into the field. The recently observed Gravitational Waves (GW) GW170817 \cite{PhysRevLett.119.161101,PhysRevLett.121.161101} and GW190425 \cite{Abbott_2020} have constrained the NS maximum mass and radius. The presence of exotic phases like quarks inside NS has also been observed recently \cite{Annala}. After the discovery of GW170817, more theoretical work has been done to understand the relation between EoS and NS properties through various aspects like a phase transition, symmetry energy \cite{PhysRevLett.120.172703,PhysRevD.97.084038,PhysRevC.98.035804,PhysRevC.97.045806,PhysRevLett.120.172702,PhysRevLett.120.261103,PhysRevLett.121.062701,iarather,Rather_2020}. However, there are still numerous fundamental problems related to the matter under extreme conditions that are yet to be answered. The most important one is the unified model which can describe the overall structure of a NS, from the outer crust to the inner core.\par
The structure and the properties of a NS are determined by the Equation of State (EoS) which describes the state of matter under given physical conditions. The discovery of massive NS's provide constrain on NS maximum mass and radius \cite{Antoniadis1233232,Fonseca_2016,2010Natur.467.1081D}. To determine the pressure and the energy density of a NS matter, the isospin asymmetry and the binding energy act as the key inputs.  The symmetry energy $S(\rho)$ and its derivatives ($L_{sym}$,$K_{sym}$, $Q_{sym}$) affect the EoS and have been constrained in the narrow limits through the finite nuclei properties \cite{MORFOUACE2019135045,Lattimer_2013,agrawal2020constraining}. For the density-dependent symmetry energy, a strong correlation is found between the pressure at saturation density and the radius of a NS \cite{2001ApJ...550..426L}. It has been shown \cite{PhysRevLett.85.5296,PhysRevC.64.027302} that the slope parameter $L_{sym}$ is strongly correlated to the neutron skin thickness of heavy nuclei like $^{208}Pb$. A higher $L_{sym}$ value describes neutron matter with higher pressure and hence thicker neutron skin \cite{PhysRevLett.106.252501,FURNSTAHL200285,PhysRevLett.86.5647}. The curvature parameter $K_{sym}$ determines the crust-core transition density and the gravitational binding energy of the NS \cite{TSANG2019533}. The skewness parameter $Q_{sym}$ has a large saturation in its value due to different values from various models. The skewness parameter is related to the nuclear incompressibility of the system \cite{typel}. All these quantities affect the EoS directly or indirectly. \par
The unified EoS describes the NS from its outer crust to the inner core. However, a unified EoS is generally not available. Hence the complete EoS is divided into three different phases: the outer crust phase, the inner crust, and the core phase. It has been shown \cite{PhysRevC.94.035804,PhysRevC.94.015808} that the NS properties like mass and radius do not depend upon the outer crust EoS, but a particular choice of inner crust EoS and the matching of inner crust to the core EoS is critical and the variations larger than 0.5 km have been obtained for a 1.4 M$_{\odot}$. For the outer crust which lies in the density range 10$^4$-10$^{11}$ g/cm$^{3}$, the Baym-Pethick-Sutherland (BPS) EoS \cite{Baym:1971pw}, the Haensel-Pichon (HP) EoS \cite{1994A&A...283..313H} are widely used in the literature. Both these EoSs do not affect the mass and radius of a NS. For the matter beyond the neutron drip density (4$\times$10$^{11}$ g/cm$^{3}$) ranging from density 10$^{11}$-10$^{14}$ g/cm$^{3}$, the inner crust EoS follows. The Baym-Bethe-Pethick (BBP) EoS \cite{BAYM1971225} is usually used. However to avoid the large uncertainties in mass and radius of a NS, studies have shown that for the complete unified EoS, the inner crust EoS should be either from the same model as core EoS or the symmetry energy slope parameter should match \cite{PhysRevC.94.015808,PhysRevC.90.045803}.  \par
The observation of the mass and the radius of static as well as rotating NSs may provide some useful constraints on the EoS. The NSs, due to their compactness, can rotate very fast as compared to the other astrophysical observables \cite{friedman_stergioulas_2013}. Hence the measurements of rotating neutron star properties like mass, radius, frequency, moment of inertia etc. close to the mass-shielding limit may lead to more constraints on the composition of the nuclear matter at higher densities and the EoS.  Rotating NSs proivde more information about the structure and the composition of the star through the measurement of more quantities than the non-rotating ones. Quantities like moment of inertia, eccentricity provide the information about how fast an object can spin with a given angular momentum and the deformation of the mass while spinning. The universal relations between these bulk properties of NSs may help to impose constraints on the radius of a NS. The choice of inner crust EoS on the static and rns will affect the mass and radius and consequently the other star matter properties.\par 
This paper is organized as follows. In section \ref{sec:1}, we explain the relativistic-mean field formalism to the study of the EoS. The Thomas-Fermi approximation within RMF framework to describe the inner crust part and the transition from crust to the core is also discussed. Section \ref{sec:2} discusses the nuclear matter (NM) properties for parameter sets used.  The unified EoS by combining the outer crust, inner crust, and the core EoS  will  be discussed in section \ref{sec:2}. The NS properties for static as well as rotating NS's with different EoSs are also discussed. Finally, the summary and conclusions are outlined in section \ref{sec:3}.

\section{Theory and Formalism}
\label{sec:1}
In this section, we summarize the formulations of our EoS for NM. The Relativistic Mean Field (RMF) theory in which the nucleons interact through the meson exchange is adopted. The simplest relativistic Lagrangian contains the contribution from $\sigma$, $\omega$, and $\rho$ mesons without any cross-coupling between them \cite{walecka}. The prediction of large NM incompressibility \cite{1974AnPhy..83..491W} by this model was reduced to an acceptable range by the addition of self-coupling terms \cite{bodmer}. The addition of other self- and cross-couplings improved the NM properties. The Effective field theory motivated RMF (E-RMF) is an extension to the basic RMF theory in which all possible self- and cross-couplings between the mesons are included \cite{FURNSTAHL1996539,FURNSTAHL1997441,bharat}. The E-RMF Lagrangian used in the present work,  which contains the contribution from $\sigma$-, $\omega$- mesons upto 4th order expansion, and  $\rho$- and $\delta$- mesons upto 2nd order is given by \cite{PhysRevC.89.044001}

\begin{eqnarray}\label{eq.1}
\mathcal{L} = \sum_{\alpha =n,p} \bar{\psi}_{\alpha} \Biggl\{\gamma_{\mu}(i\partial^{\mu}-g_{\omega} \omega^{\mu} - \frac{1}{2}g_{\rho} \bm{\tau}_{\alpha}\cdot \bm{\rho}^{\mu} )-(M-g_{\sigma}\sigma) 
-g_{\delta}\bm{\tau}_{\alpha}\cdot \bm{\delta} \Biggr\} \psi_{\alpha} \nonumber \\ 
+\frac{1}{2}\partial^{\mu}\sigma \partial_{\mu}\sigma-\frac{1}{2}m_{\sigma}^2\sigma^2 +\frac{\zeta_0}{4!}g_{\omega}^2 (\omega^{\mu}\omega_{\mu})^2  -g_{\sigma} \frac{m_{\sigma}^2}{M} \Bigg(\frac{k_3}{3!}+\frac{k_4}{4!}\frac{g_{\sigma}}{M}\sigma\Bigg)\sigma^3 \nonumber \\
+\frac{1}{2}m_{\omega}^2 \omega^{\mu}\omega_{\mu} -\frac{1}{4} {\textbf{F}}^{\mu \nu} \cdot {\textbf{F}}_{\mu \nu}  +\frac{1}{2}\frac{g_{\sigma}\sigma}{M}\Bigg(\eta_1+\frac{\eta_2}{2}\frac{g_{\sigma}}{M}\sigma\Bigg)m_{\omega}^2 \omega^{\mu}\omega_{\mu} \nonumber \\
+\frac{1}{2}\eta_{\rho}\frac{m_{\rho}^2}{M}
g_{\sigma}\sigma (\bm{\rho}^{\mu}\cdot \bm{\rho}_{\mu}) 
+\frac{1}{2}m_{\rho}^2 \rho^{\mu}\cdot \rho_{\mu} -\frac{1}{4} {\textbf{R}}^{\mu \nu}\cdot {\textbf{R}}_{\mu \nu}  \nonumber \\
-\Lambda_{\omega} g_{\omega}^2 g_{\rho}^2(\omega^{\mu}\omega_{\mu})(\bm{\rho}^{\mu}\cdot \bm{\rho}_{\mu})  +\frac{1}{2}\partial^{\mu}\bm{\delta} \partial_{\mu}\bm{\delta}-\frac{1}{2} m_{\delta}^2\bm{\delta}^2 ,	
\end{eqnarray}

where $\psi$ is the nucleonic field and $M$ is the nucleonic mass. $m_{\sigma}$, $m_{\omega}$,$m_{\rho}$, and $m_{\delta}$ are the masses and  $g_{\sigma}$, $g_{\omega}$,$g_{\rho}$, and $g_{\delta}$ are the coupling constants of $\sigma$, $\omega$, $\rho$, and $\delta$ mesons respectively. The Euler-Lagrangian equations of motion for the meson fields are obtained using the relativistic mean field approximation \cite{PhysRevC.97.045806}. 
\begin{eqnarray}
m_{\sigma}^2 \sigma  =g_{\sigma} \rho_s(r)  - \frac{m_{\sigma}^2 g_{\sigma}}{M} \sigma^2 \Bigg(\frac{k_3}{2}+\frac{k_4}{6}\frac{g_{\sigma}\sigma}{M}\Bigg) \nonumber \\
+ \frac{g_{\sigma}}{2M} \Bigg(\eta_1+\eta_2\frac{g_{\sigma}\sigma}{M}\Bigg) m_{\omega}^2 \omega^2+\frac{\eta_{\rho}}{2M}g_{\sigma} m_{\omega}^2 \rho^2,
\end{eqnarray}

\begin{eqnarray}
m_{\omega}^2 \omega  =g_{\omega} \rho(r) - \frac{g_{\sigma}}{M} \Bigg(\eta_1+\eta_2\frac{g_{\sigma}\sigma}{M}\Bigg) m_{\omega}^2 \omega -\frac{1}{6}\zeta_0 g_{\omega}^2 \omega^3 \nonumber\\
- 2\Lambda_{\omega} (g_{\omega} g_{\rho} \rho)^2  \omega, 
\end{eqnarray}

\begin{eqnarray}
m_{\rho}^2 \rho = \frac{1}{2}g_{\rho} \rho_3 (r)-\eta_{\rho} \frac{g_{\sigma}\sigma}{M} m_{\rho}^2 \rho  - 2\Lambda_{\omega} (g_{\omega} g_{\rho} \omega)^2  \rho,
\end{eqnarray}

\begin{equation}
m_{\delta}^2 \delta =g_{\delta} \rho_{s3}(r),
\end{equation}

where,

\begin{eqnarray}
\rho_s(r) = \sum_{\alpha =n,p} <\bar{\psi}_{\alpha}\gamma_0 \psi_{\alpha}> = \rho_{sn} +\rho_{sp}  \nonumber \\
= \frac{2}{(2\pi)^3} \sum_{\alpha} \Bigg(\int  \frac{M^*_{\alpha}}{(k_{\alpha}^2 +M_{\alpha}^{*2})^{1/2}} d^3k\Bigg),
\end{eqnarray}

\begin{eqnarray}
\rho(r) = \sum_{\alpha} <\bar{\psi}_{\alpha} \psi_{\alpha}> 
=\rho_n +\rho_p   \nonumber \\
= \frac{2}{(2\pi)^3} \sum_{\alpha} \int d^3k ,
\end{eqnarray}

\begin{eqnarray}
\rho_3(r)  = \sum_{\alpha} <\bar{\psi}_{\alpha} \tau_3 \psi_{\alpha}> = \rho_p -\rho_n,
\end{eqnarray}
and
\begin{eqnarray}
\rho_{s3}(r)  = \sum_{\alpha} <\bar{\psi}_{\alpha} \tau_3 \gamma_0 \psi_{\alpha}> = \rho_{sp} -\rho_{sn}.
\end{eqnarray}

are the scalar, baryon, and isovector densities respectively.  $M_{\alpha}^*  (\alpha=n,p)$ is the effective mass of nucleons given by the relation
\begin{eqnarray}
M_n^* = M-g_{\sigma} \sigma +g_{\delta} \delta, \\
M_p^* = M-g_{\sigma} \sigma -g_{\delta} \delta.
\end{eqnarray}

The expression for the energy density and pressure are obtained from the given Lagrangian using energy momentum tensor relation given by \cite{Serot:1984ey}
\begin{equation}
T^{\mu \nu} = \sum_i \partial_{\nu} \phi_i \frac{\partial \mathcal {L}}{\partial(\partial^{\mu} \phi_i)} -g_{\mu \nu} \mathcal{L}.
\end{equation} 

The energy density follows from  the zeroth component and the pressure from the third component of the energy momentum tensor. Their expressions are given as \cite{PhysRevC.89.044001}

\begin{eqnarray}
\mathcal{E} = \frac{2}{(2\pi)^3}\sum_{\alpha =n,p}\int_{0}^{k_{\alpha}}d^3k E_i^*(k) + \rho g_{\omega}\omega 
+m_{\sigma}^2 \sigma^2 \Bigg(\frac{1}{2}+\frac{k_3}{3!}\frac{g_{\sigma}\sigma}{M}+\frac{k_4}{4!}\frac{g_{\sigma}^2\sigma^2}{M^2}\Bigg) \nonumber \\ 
- \frac{1}{4!}\zeta_0 g_{\omega}^2 \omega^4 
-\frac{1}{2}m_{\omega}^2 \omega^2 \Bigg(1+\eta_1\frac{g_{\sigma}\sigma}{M}+\frac{\eta_2}{2}\frac{g_{\sigma}^2\sigma^2}{M^2}\Bigg) 
+\frac{1}{2}\rho_3 g_{\rho}\rho \nonumber \\ 
-\frac{1}{2}\Bigg(1+\frac{\eta_{\rho}g_{\sigma}\sigma}{M}\Bigg)m_{\rho}^2 \rho^2 
-\Lambda_{\omega}g_{\rho}^2 g_{\omega}^2 \rho^2 \omega^2 +\frac{1}{2}m_{\delta}^2 \delta^2 , 
\end{eqnarray}

and

\begin{eqnarray}
P= \frac{2}{3(2\pi)^3}\sum_{\alpha =n,p}\int_{0}^{k_{\alpha}}d^3k \frac{k^2}{E_i^*(k)} 
- m_{\sigma}^2 \sigma^2 \Bigg(\frac{1}{2}+\frac{k_3}{3!}\frac{g_{\sigma}\sigma}{M}+\frac{k_4}{4!}\frac{g_{\sigma}^2\sigma^2}{M^2}\Bigg) 
+ \frac{1}{4!}\zeta_0 g_{\omega}^2 \omega^4 \nonumber \\
+\frac{1}{2}m_{\omega}^2 \omega^2 \Bigg(1+\eta_1\frac{g_{\sigma}\sigma}{M}+\frac{\eta_2}{2}\frac{g_{\sigma}^2\sigma^2}{M^2}\Bigg)  +\frac{1}{2}\rho_3 g_{\rho}\rho \nonumber  \\
+\frac{1}{2}\Bigg(1+\frac{\eta_{\rho}g_{\sigma}\sigma}{M}\Bigg)m_{\rho}^2 \rho^2 
+\Lambda_{\omega}g_{\rho}^2 g_{\omega}^2 \rho^2 \omega^2 -\frac{1}{2}m_{\delta}^2 \delta^2.
\end{eqnarray}

The symmetry energy usually defined as the difference in the energy per nucleon of pure neutron matter (PNM) and symmetric NM (SNM) plays a major role in the nuclear EoS \cite{PhysRevC.93.035806}. The symmetry energy and its derivatives have a major impact on NS properties like mass and radius. The symmetry energy is defined as

\begin{equation}
S(\rho) = \frac{1}{2} \Bigg(\frac{\partial^2 \mathcal{E}(\rho,\beta)}{\partial \beta^2}\Bigg)_{\beta=0}
\end{equation}

where, $\beta$ is the asymmetric parameter which is 0 for SNM and 1 for PNM.

\begin{equation}
\beta=\frac{\rho_n -\rho_p}{\rho_n +\rho_p}
\end{equation}

The derivatives of symmetry energy, slope parameter $L_{sym}$, symmetry energy curvature $K_{sym}$, and the skewness parameter $Q_{sym}$ are defined at the saturation density $\rho_0$ as

\begin{equation}
L_{sym}= 3\rho \frac{\partial S(\rho)}{\partial \rho}\Bigg|_{\rho=\rho_0},
\end{equation} 
\begin{equation}
K_{sym}= 9\rho^2 \frac{\partial^2 S(\rho)}{\partial \rho^2}\Bigg|_{\rho=\rho_0},
\end{equation}
and \\

\begin{equation}
Q_{sym}= 27\rho^3 \frac{\partial^3 S(\rho)}{\partial \rho^3}\Bigg|_{\rho=\rho_0},
\end{equation} 
respectively.\par 
The inner crust which contains the inhomogenous NM is defined by applying the Thomas-Fermi(TF) approximation \cite{10.1143/PTP.100.1013,SHEN1998435} within the same RMF model \cite{PhysRevC.72.015802,PhysRevC.78.015802,PhysRevC.79.035804,PhysRevC.82.055807}. The Skyrme type interactions like Lattimer and Swesty EoS \cite{LATTIMER1991331} and the compressible liquid drop model have been widely used in the literature by several authors to describe the nonuniform matter \cite{Abbott_2017,CHABANAT1997710,Chamel2008}. \par 
The outer crust of a NS consists of nuclei distributed in a solid body-centered-cubic (bcc) lattice filled by free electron gas. Moving from outer crust to the inner crust part, the increasing density leads to the complete ionization of all the atoms and beyond a certain density, leads to the formation of a quasiuniform gas. As we move to the greater density regions, more and more electrons are captured by the nuclei, which thus become neutron-rich. At the density around 4.2$\times$10$^{11}$ g cm$^{-3}$, neutron drip sets in. With the help of nuclear models, the neutron drip density determines the boundary between the outer crust and the inner crust. The inner crust part contains the free electrons and the neutron gases, forming different types of pasta structures. Both the inner and the outer crust densities are a fraction of the normal nuclear density.  \par 
The transition density at the crust-core interface is uncertain due to the insufficient knowledge of the neutron-rich NM EoS. Furthermore, the determination of the transition density is very complicated as the inner crust part may contain some internal structures (pasta phases). A well defined approach is to find the density at which the uniform liquid phase becomes unstable against small density fluctuations, indicating the formation of  nuclear clusters. This boundary between the inhomogenous solid crust and the liquid core is connected to the isospin dependence of nuclear models below the saturation value by the widely used thermodynamic method \cite{PhysRevC.70.065804,PhysRevC.76.025801,PhysRevC.86.015801,PhysRevC.89.028801,Routray_2016}. In the present work, we use the thermodynamic method to determine the crust-core transition boundary, which allows the stability of the NS's to be determined in terms of its bulk properties. The following conditions determine the boundary between the crust and the core:
\begin{equation}\label{e1}
-\Bigg(\frac{\partial P}{\partial v}\Bigg)_{\mu} > 0,
\end{equation} 

\begin{equation}\label{e2}
-\Bigg(\frac{\partial \mu_{np}}{\partial q}\Bigg)_{v} > 0.
\end{equation}
Here $P$ is the pressure of the $\beta$-stable matter, $\mu_{np}$ is the difference between neutron and proton chemical potentials, $v=1/\rho$ is the volume per baryon, and $q$ is the charge per baryon. The stability conditions in equations (\ref{e1}) and (\ref{e2}) refer to the mechanical and chemical stabilities of the system respectively. The total pressure $P$ in equation (\ref{e1}) contains the contribution from baryonic part and the leptonic part. The leptonic contribution to the pressure becomes a function of chemical potential $\mu$ only when the system is in beta-equilibrium and hence does not contribute to the stability condition in equation (\ref{e1}), which now becomes
\begin{equation}\label{e3}
-\Bigg(\frac{\partial P_N}{\partial v}\Bigg)_{\mu} > 0.
\end{equation} 
The charge fraction $q= Y_p -Y_e = Y_p -\rho_e/\rho$, where $\rho_e = \mu_e^3/(3\pi^2)$ follows that
\begin{equation}\label{e4}
-\Bigg(\frac{\partial q}{\partial \mu}\Bigg)_{v} = -\Bigg(\frac{\partial Y_p}{\partial \mu}\Bigg)_{v} +\frac{1}{\rho} \Bigg(\frac{\partial \rho_e}{\partial \mu}\Bigg)_{v}.
\end{equation} 
With $v=1/\rho$ and $\mu=-\partial e(\rho, Y_p)/\partial Y_p$, the equations (\ref{e3}) and (\ref{e4}) reduce to
\begin{equation}\label{e5}
\rho^2\Bigg(\frac{\partial P_N}{\partial \rho}\Bigg)_{\mu} = 2\rho^3\Bigg(\frac{\partial e(\rho,Y_p)}{\partial \rho}\Bigg) +\rho^4 \Bigg(\frac{\partial^2 e(\rho, Y_p)}{\partial \rho^2}\Bigg)-\rho^4 \frac{\Big(\frac{\partial^2 e(\rho,Y_p)}{\partial \rho \partial Y_p}\Big)^2}{\Big(\frac{\partial^2 e(\rho,Y_p)}{\partial Y_p^2}\Big)} >0,
\end{equation}

\begin{equation}\label{e6}
-\Bigg(\frac{\partial q}{\partial \mu}\Bigg)_{v} = \Bigg(\frac{\partial^2 e(\rho,Y_p)}{\partial Y_p^2}\Bigg)^{-1} + \frac{\mu^2}{\pi^2 \rho}>0.
\end{equation}
The first two terms on the rhs of eq. (\ref{e5}) refer to the nuclear pressure and NM incompressibility which are positive. The third term comes from the leptonic part and contributes negatively.  The second term on the rhs of eq.(\ref{e6}) can be recognized as the screening length, while the first term on rhs is proportional to the nuclear symmetry energy $E_{sym}(\rho)$ under parabolic approximation. The condition in eq.(\ref{e6}) is satisfied as the heavy-ion collision studies favor an increasing behaviour of  $E_{sym}(\rho)$. The stability condition in eq.(\ref{e5}) thus reduces to
\begin{equation}\label{e7}
V_{thermal} = 2\rho\Bigg(\frac{\partial e(\rho,Y_p)}{\partial \rho}\Bigg) +\rho^2\Bigg(\frac{\partial^2 e(\rho, Y_p)}{\partial \rho^2}\Bigg)- \frac{\Big(\frac{\partial^2 e(\rho,Y_p)}{\partial \rho \partial Y_p}\Big)^2}{\Big(\frac{\partial^2 e(\rho,Y_p)}{\partial Y_p^2}\Big)} >0,
\end{equation}
where $V_{thermal}=\Big(\frac{\partial P_N}{\partial \rho}\Big)_{\mu}$ and $e(\rho,Y_p) =H(\rho, Y_p)/\rho$ is the energy per particle of asymmetric  nuclear matter (ANM) with $H(\rho, Y_p)$ as the energy density. From the above relation, the determination of the transition density for the crust-core interface using the EoS of ANM is evident. For a given nucleonic density $\rho$, once the proton fraction is ascertained from the charge neutral and beta stability conditions, the nucleonic energy per particle can be calculated and hence the expression (\ref{e7}) can be evaluated. The stability condition of $V_{thermal}$ is a signature of the crust-core transition in the beta-stable NS matter. A more detailed description of the crust-core transition is given in ref's \cite{PhysRevC.86.015801,PhysRevC.76.025801,Routray_2016}.
\section{Results and Discussion}
\label{sec:2}
To study the effect of crust EoS on NS properties, we have chosen several parameter sets to construct the core EoS. Since the outer crust EoS does not affect the NS maximum mass and radius, therefore the BPS EoS \cite{Baym:1971pw} has been used for the outer crust part. For the inner crust part, the relativistic mean-field model with constant couplings, non-linear terms \cite{nonlinear}, and density-dependent couplings \cite{Typel:1999yq} have been used. These include NL3 \cite{PhysRevC.55.540} set with non-linear $\sigma$ terms, TM1 \cite{Tm1} with non-linear $\sigma$ and $\omega$ terms, NL3$\omega \rho$ \cite{PhysRevC.64.062802,PhysRevLett.86.5647} which includes the non-linear $\omega \rho$ terms in addition to the previous couplings, FSU \cite{PhysRevLett.95.122501} and IU-FSU \cite{PhysRevC.82.055803}, and the density-dependent DD-ME2 \cite{PhysRevC.71.024312} and DD-ME$\delta$ \cite{PhysRevC.84.054309}.\par
The NM properties at saturation density for the above considered crust EoSs are shown in table \ref{tbl1}. The symmetry energy slope parameter $L_{sym}$ of the given sets lies in the range 47-118 MeV. We have considered these sets for the inner crust part to determine the variation in the maximum mass and the corresponding radius of a NS.

\begin{table}[ht]
	\centering
	\caption{NM properties for the inner crust EoS at saturation density ($\rho_0$), energy ($E_0$), symmetry energy ($S$), slope parameter ($L_{sym}$), incompressibility coefficient ($K$), and skewness parameter ($Q_{sym}$). All the values are in MeV except for the ($\rho_0$) which is in fm$^{-3}$.}
	{\begin{tabular}{ccccccc}
			\hline 
			Model & $\rho_0$ & $E_0$ & $S$ & $L_{sym}$ & $K$ & $Q_{sym}$\\
			\hline
			NL3 & 0.148 & -16.24 & 37.3 & 118.3 & 270.7 & 203 \\
			TM1 & 0.145 & -16.26 & 36.8 & 110.6 & 280.4 & -295 \\
			FSU & 0.148 & -16.30 & 32.6 & 60.5 & 230.0 & -523 \\
			IU-FSU & 0.155 & -16.40 & 31.3 & 47.2 & 213.2 & -288 \\
			NL3$\omega \rho$ & 0.148 & -16.30 & 31.7 & 55.2 & 272.0 & 203 \\
			DD-ME2 & 0.152 & -16.14 & 32.3 & 51.4 & 250.8 & 478 \\
			DD-ME$\delta$ & 0.152 & -16.12 & 32.4 & 52.9 & 219.1 & -741 \\
			\hline
		\end{tabular}\label{tbl1}}
\end{table}

The E-RMF formalism is used to construct the core EoS.  We covered a wide range of models containing only $\sigma$ self-coupling terms to the models with all types of self- and cross-couplings along with the $\delta$ meson inclusion.  NL3 \cite{PhysRevC.55.540}, TM1 \cite{Tm1}, IU-FSU \cite{PhysRevC.100.025805}, IOPB-I \cite{PhysRevC.97.045806}, and G3 \cite{bharat} parameter sets are used to study the NS core. All the coupling constants and the meson masses for the above parameter sets are given in \cite{Tm1,PhysRevC.82.055803,PhysRevC.97.045806}. The parameter sets for the core EoS used in the present study cover a NS maximum mass range from $\approx$2-2.8$M_{\odot}$. This allows us to determine the variation in the properties of the NS for a range of maximum mass with the symmetry energy. \par
\begin{figure}[ht]
	\centering
	\includegraphics[width=10cm, height=8cm]{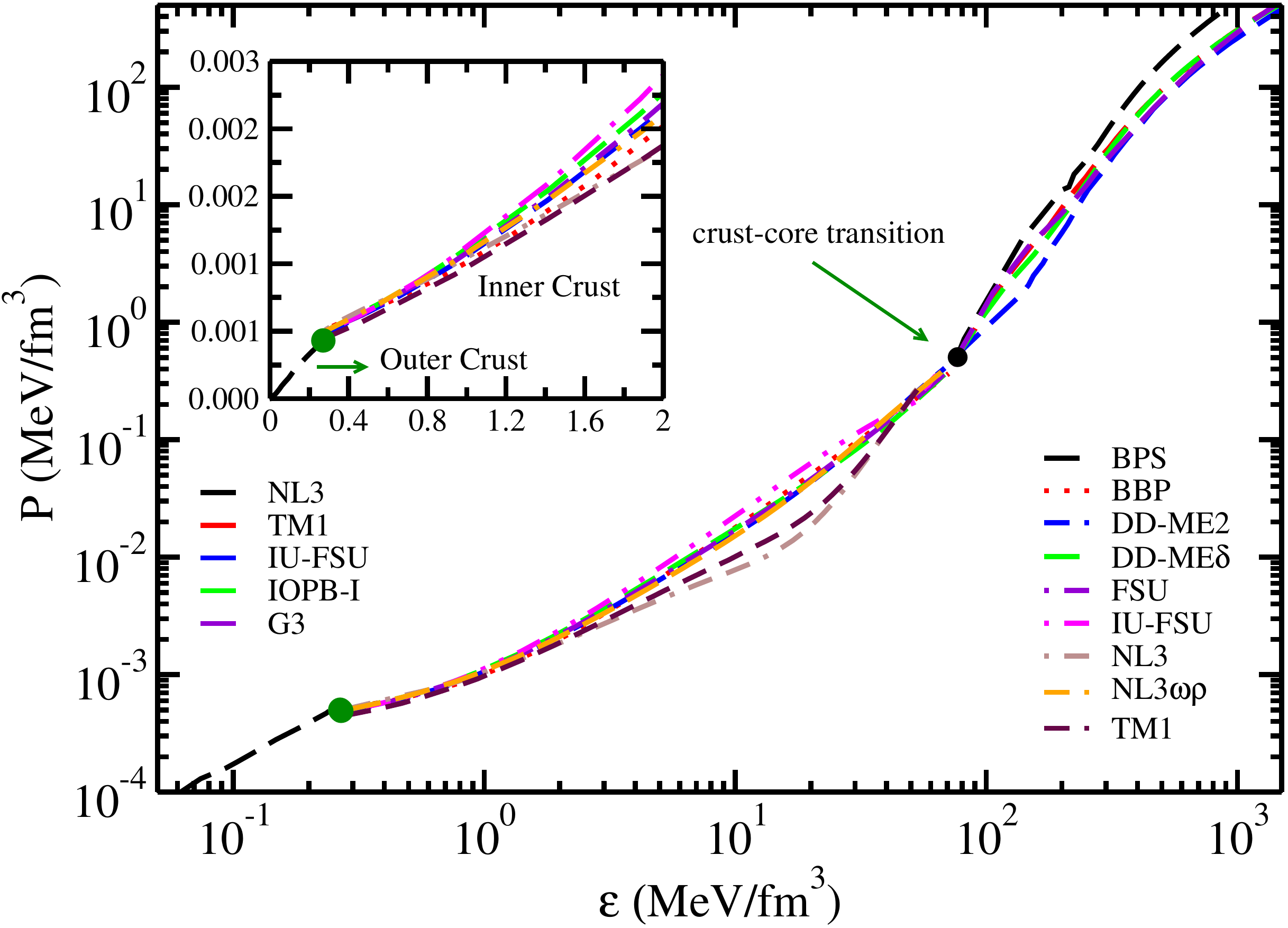}
	\caption{Unified EoS with different inner crust and core EoS. The inset shows the matching of outer crust with the inner crust. The green dot shows the matching point of BPS EoS with the inner crust. The black dot represents the point of crust-core transition.}
	\label{FIG:1}
\end{figure}

The complete EoS consisting of the outer crust, the inner crust, and the core are constructed using the above defined parameter sets. The unified and non-unified EoSs follow as: BPS (for the outer crust)+ BBP, NL3, TM1, NL3$\omega \rho$, FSU, IU-FSU, DD-ME2, and DD-ME$\delta$ (for the inner crust)+ NL3, TM1, IU-FSU, IOPB-I, and G3 (for the core). The EoS without inner crust is also constructed to see the impact of inner crust on NS properties. All the inner crust EoSs used are taken from the reference \cite{PhysRevC.90.045803}. The different unified and non-unified EoSs produced are shown in figure \ref{FIG:1}.

The green dot in the inset (as well as in the main plot) of figure \ref{FIG:1} shows the matching of outer crust with inner crust EoS while the black dot in the main plot represents the crust-core transition point.
The NL3 parameter set produces stiff core EoS as compared to other parameter sets. IU-FSU produces soft EoS at low density. G3 set produces soft EoS as compared to NL3, TM1 and IOPB-I at both low and high energy densities. Among the inner crust EoS, NL3 and TM1 set produce soft EoS at very low density and become stiff as the density increases. IU-FSU crust initially produces stiff EoS but becomes soft at higher energy density ($\mathcal{E}$ $\approx$ 45 MeV/fm$^3$). For outer crust, the BPS EoS is used for all different combinations of inner crust and core EoSs as the outer crust EoS doesn't affect the mass and radius of a NS. \par
To determine the maximum mass and the corresponding radius of a stationary and spherical NS's obtained using different EoSs, we use the Tolman-Oppenheimer-Volkoff (TOV) equations \cite{PhysRev.55.374,PhysRev.55.364}.

\begin{equation}\label{tov1}
\frac{dP(r)}{dr}= -\frac{[\mathcal{E}(r) +P(r)][M(r)+4\pi r^3 P(r)]}{r^2(1-2M(r)/r) },
\end{equation}
and
\begin{equation}\label{tov2}
\frac{dM(r)}{dr}= 4\pi r^2 \mathcal{E}(r).
\end{equation}

Here $M(r)$ is the gravitational mass. For the given boundary conditions $P(0)=P_c$, $M(0)=0$, with $P_c$ being the central pressure, the equations (\ref{tov1}) and (\ref{tov2}) are solved to obtain the NS properties. Apart from obtaining the properties of a static NS, we also see the impact of inner crust on maximally rotating NS (rns). The properties of rotating NS are obtained using the RNS code  \cite{Stergioulas2003,Stergioulas_1995,rnscode,refId0}. \par

\begin{figure}[hbt!]
	\centering
	\includegraphics[width=10cm, height=8cm]{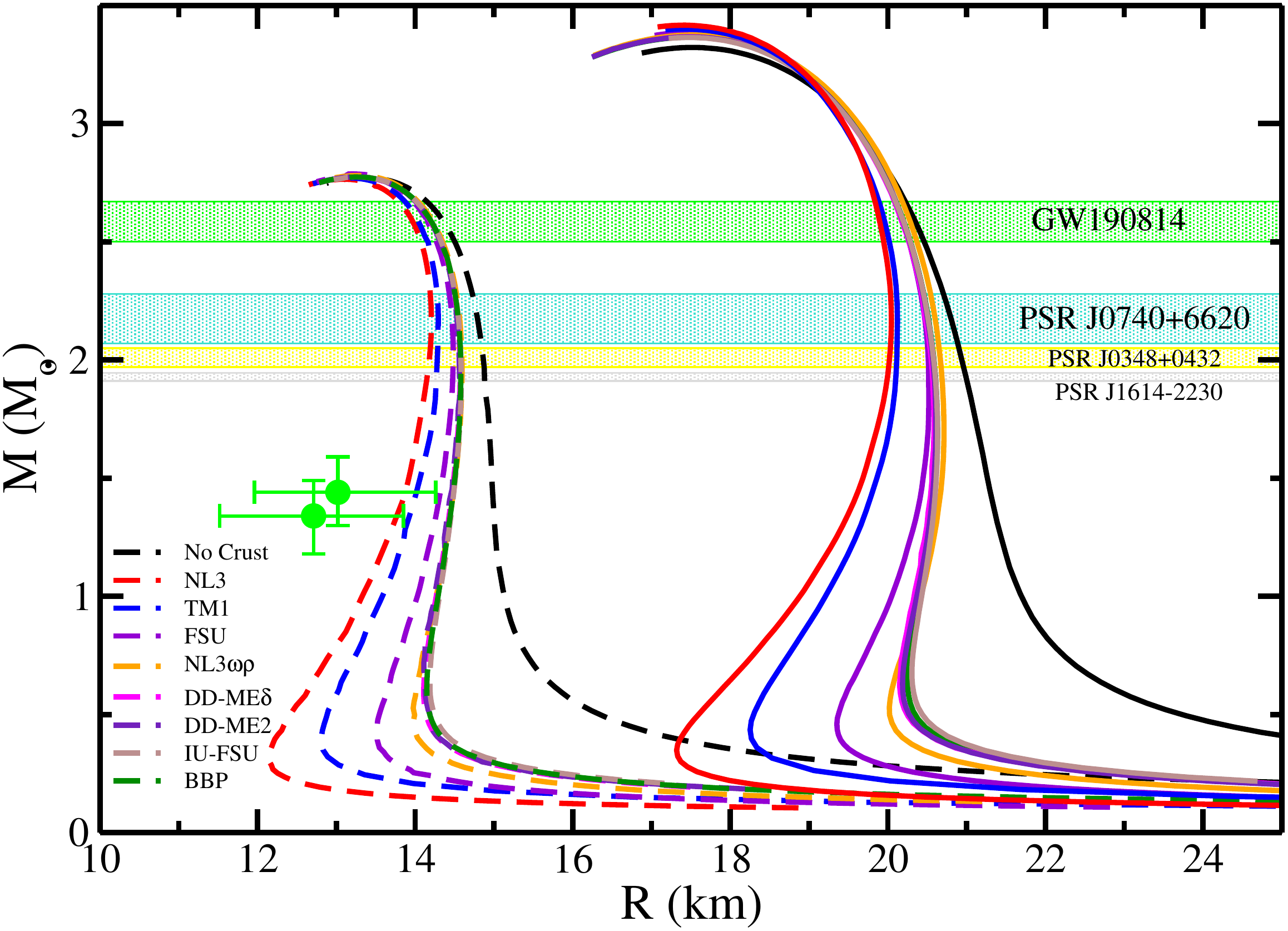}
	\caption{Mass-Radius profile of SNS and RNS for NL3 core with different inner crust EoS. The solid (dashed) lines represent the rotating NS's (static NS's) The recent constraints on the mass \protect\cite{2010Natur.467.1081D,Antoniadis1233232,Cromartie2020,Abbott_2020a} and radius from NICER measurements \protect\cite{Riley_2019,Miller_2019} of NS are also shown.}
	\label{FIG:2}
\end{figure}

Figure \ref{FIG:2} shows the mass-radius relation for a static and maximally rotating NS with NL3 core with different inner crust EoS. The shaded regions represent the constraints on the maximum mass of a NS by pulsars PSR J1614-2230 (1.928$\pm$0.017)$M_{\odot}$ \cite{2010Natur.467.1081D}, PSR J0348+0432(2.01$\pm$0.04)$M_{\odot}$ \cite{Antoniadis1233232}, and PSR J0740+6620 (2.04$^{+0.10}_{-0.09}$)$M_{\odot}$ \cite{Cromartie2020},  and GW190814 (2.50-2.67 $M_{\odot}$) \cite{Abbott_2020a}. The NICER constraints on the stellar radius obtained from the measurement of millisecond pulsar (MSP) PSR J0030+0451 at the inferred mass $M=1.34_{-0.16}^{+0.15} M_{\odot}$ and $M=1.44_{-0.14}^{+0.15} M_{\odot}$ given by $R=12.71_{-1.19}^{+1.14}$km and $R=13.04_{-1.06}^{+1.24}$km are also shown \cite{Riley_2019, Miller_2019}.  
The upper limit on the radius at the canonical mass of a NS is found to be $R_{1.4} \le 13.76$ km \cite{PhysRevLett.120.172702}. The constraints on the maximum mass show that the theoretical prediction of a NS maximum mass should reach the limit $\approx$ 2.0 $M_{\odot}$. But the recent observation of gravitational wave data GW190814 has a secondary component with a maximum mass in the range 2.5-2.67 $M_{\odot}$. This secondary object is considered to be either a light black-hole or a supermassive NS. \par
\vspace{1.0cm} 
From figure \ref{FIG:2}, it is clear that the NS maximum mass produced using different inner crust EoS varies by small margins and lies in the range 2.764-2.787 $M_{\odot}$. The corresponding radius varies from 13.027-13.378 km. However, the radius at canonical mass is much more affected than the radius at maximum mass. For a NS without inner crust, the radius at the canonical mass is found out to be $R_{1.4}$ =14.987 km. The large value for the radius of a NS at 1.4$M_{\odot}$ without inner crust is due to the direct transition from outer crust to the core part of the star. The neutron drip density determines the outer crust-inner crust boundary and the thermodynamic method provides the transition between the inner crust and the core. With no inner crust considered, the transition density obtained for the outer crust-core boundary affects the core EoS which results in an unexpected large value of the radius. With the addition of the inner crust, the proper measurement of the transition density leads to a true value of the NS radius  which varies from 14.496-13.853 km. The NS with a small radius at the canonical mass is produced by using NL3 as inner crust EoS which satisfies the constraints by GW170817. The NL3 set has higher value of symmetry energy slope parameter $L_{sym}$ = 118.3 MeV,but matches completely with the core EoS and hence forms a unified EoS. Thus, a unified EoS produces a NS with smaller radius at the canonical mass. The other inner crust EoSs have a smaller value of slope parameter than the NL3 set and such low value slope parameter sets like IU-FSU produces a larger radius at the canonical mass as seen in the figure. Also, the NL3 inner crust EoS satisfies the radius constraints from the NICER measurements. Thus we see that the $R_{1.4}$ has a significant relation with the slope parameter. This is consistent with the work in references \cite{PhysRevLett.120.172702,PhysRevLett.121.062701}.\par

Figure (\ref{FIG:2}) also shows the MR profile for maximally rotating stars (RNS). Similar to the Static NS (SNS), the RNS maximum mass and the corresponding radius are not much affected by the inner crust EoS, but the radius at 1.4$M_{\odot}$ varies in a similar fashion as SNS in the range R$_{1.4}$=19.5-21.4 km.  \par 
\vspace{1.0cm}


\vspace{0.5cm}
\begin{adjustbox}{center, caption={a) Same as figure \ref{FIG:2}, but for a) TM1 and b) IU-FSU core EoS.},label={tm1},nofloat=figure}
	\includegraphics[scale=0.32]{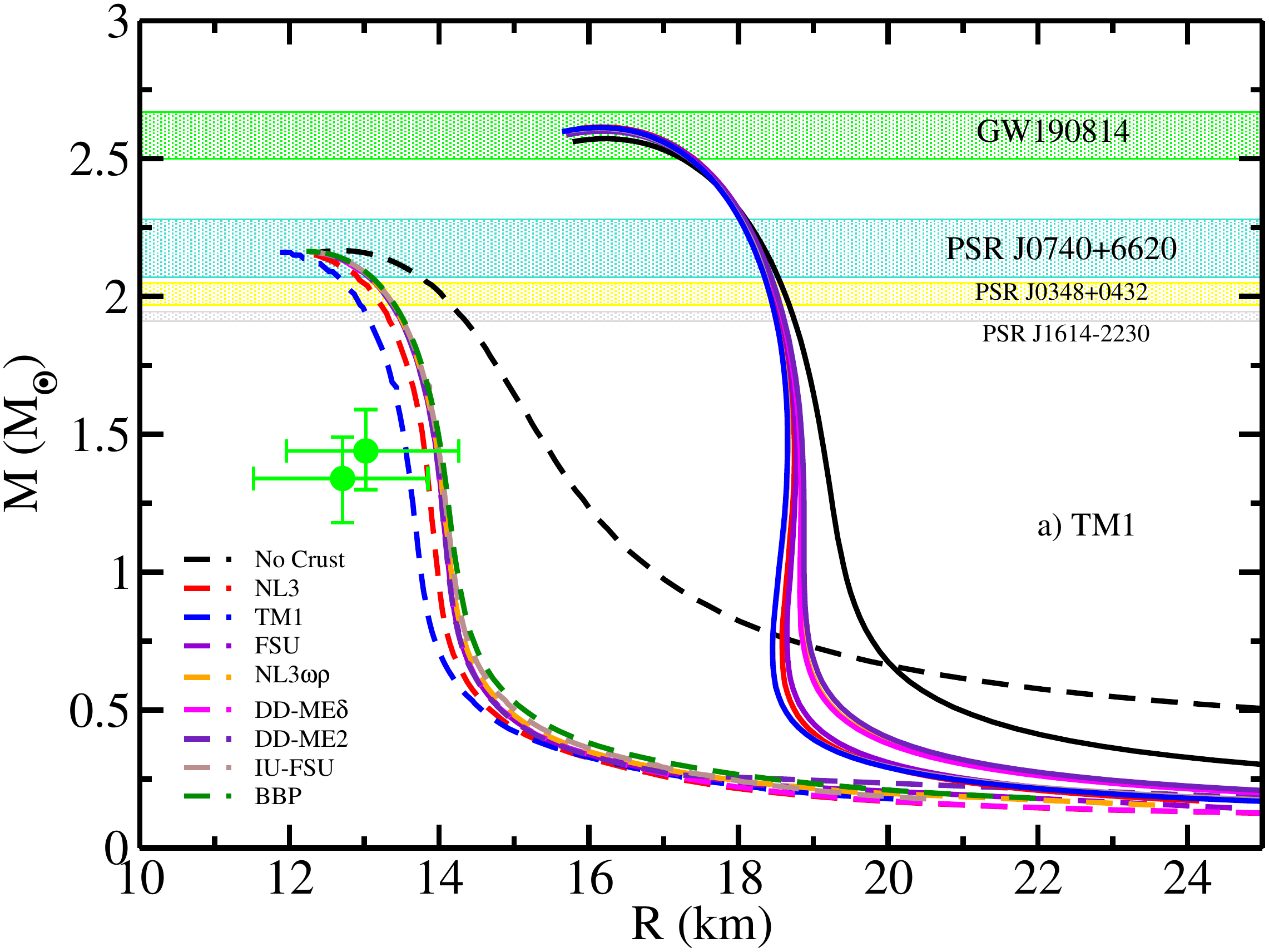}
	\hspace{0.8cm}
	\includegraphics[scale=0.32]{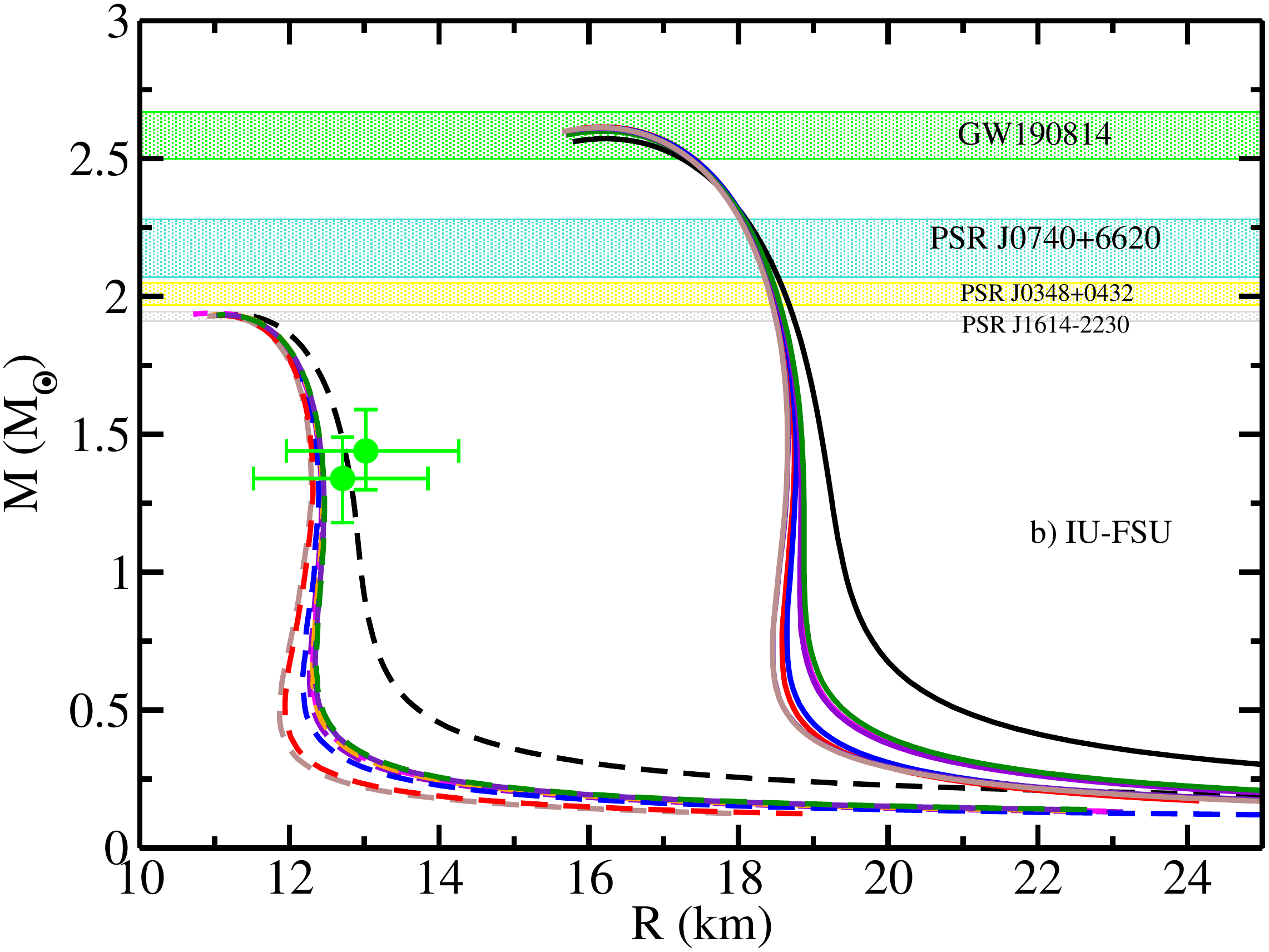}
\end{adjustbox} 

\vspace{0.5cm}

The Mass-Radius profile for TM1 and IU-FSU parameter sets are shown in figure \ref{tm1}. For TM1 core EoS with different inner crust EoS, the maximum mass and the corresponding radius lie in the range 2.141-2.177 $M_{\odot}$  and 12.234-12.805 km, respectively as shown in figure. The radius at the canonical mass varies from 13.572 km for TM1 crust EoS to 15.549 km produced without using the inner crust. As seen in the figure, although every EoS for TM1 core along with different inner crust satisfies the mass constraint from recently observed GW data, only the unified EoS (TM1 inner crust + TM1 core) satisfies the radius constraint at the canonical mass, $R_{1.4}\le 13.76$ km \cite{PhysRevLett.120.172702} . The radius of PSR J0030+0451, $R=12.71_{-1.19}^{+1.14}$ km by NICER observations \cite{Miller_2019a,Riley_2019} is also satisfied by the unified TM1 EoS. For IU-FSU core EoS, as shown in the figure, the maximum mass and radius vary from 1.931-1.940 $M_{\odot}$  and 11.030-11.263 km, respectively. The radius at the canonical mass varies from 12.295 km for IU-FSU crust EoS to 12.778 km without inner crust, all satisfying the radius constraint from NICER measurements and $R_{1.4}\le 13.76$ km. The maximum mass and the radius for RNS with TM1 and IU-FSU core are almost identical with mass in the range 2.57-2.63$M_{\odot}$ satisfying the GW190814 mass constraint and radius around 16 km. This also shows that possibility of the secondary component of GW190814 to be a maximally RNS. The unified EoS in TM1 (L$_{crust}$=110.6 MeV + L$_{core}$=110.6 MeV) and IU-FSU (L$_{crust}$=47.2 MeV + L$_{core}$=47.2 MeV) produce a NS with smaller radius for SNS as well as RNS. \par

Figure \ref{iopb} shows the MR profile for IOPB-I and G3 core EoS. For IOPB-I set, the maximum mass varies from 2.141-2.156 $M_{\odot}$ and the radius 11.872-12.029 km. $R_{1.4}$ varies from 13.118-13.508 km. 
Similar results follow for G3 EoS, where the NS maximum mass and the corresponding radius vary slightly with different inner crust EoS. However, as usual, the radius $R_{1.4}$ varies from 12.436-14.447 km. For RNS, the radius at the canonical mass varies from 18.65 to 19.18 km and 17.72 to 20.86 km for IOPB-I and G3 EoS, respectively. It is to be mentioned here that both IOPB-I anf G3 sets do not form a unified EoS i.e, inner crust and core EoS with same symmetry energy slope parameter. However, we see that for IOPB-I core EoS, the FSU inner crust with a similar value of slope parameter as IOPB-I, predicts smaller radius for NS at 1.4$M_{\odot}$ than any other crust EoS. Similar follows for G3 set ($L_{sym}$ = 49.3 MeV), the IU-FSU inner crust with a slope parameter value of 47.2 MeV gives a smaller radius NS.  \par
\vspace{1.0cm}
\begin{adjustbox}{center, caption={a) Same as figure \ref{FIG:2}, but for a) IOPB-I and b) G3 core EoS.},label={iopb},nofloat=figure}
	\includegraphics[scale=0.33]{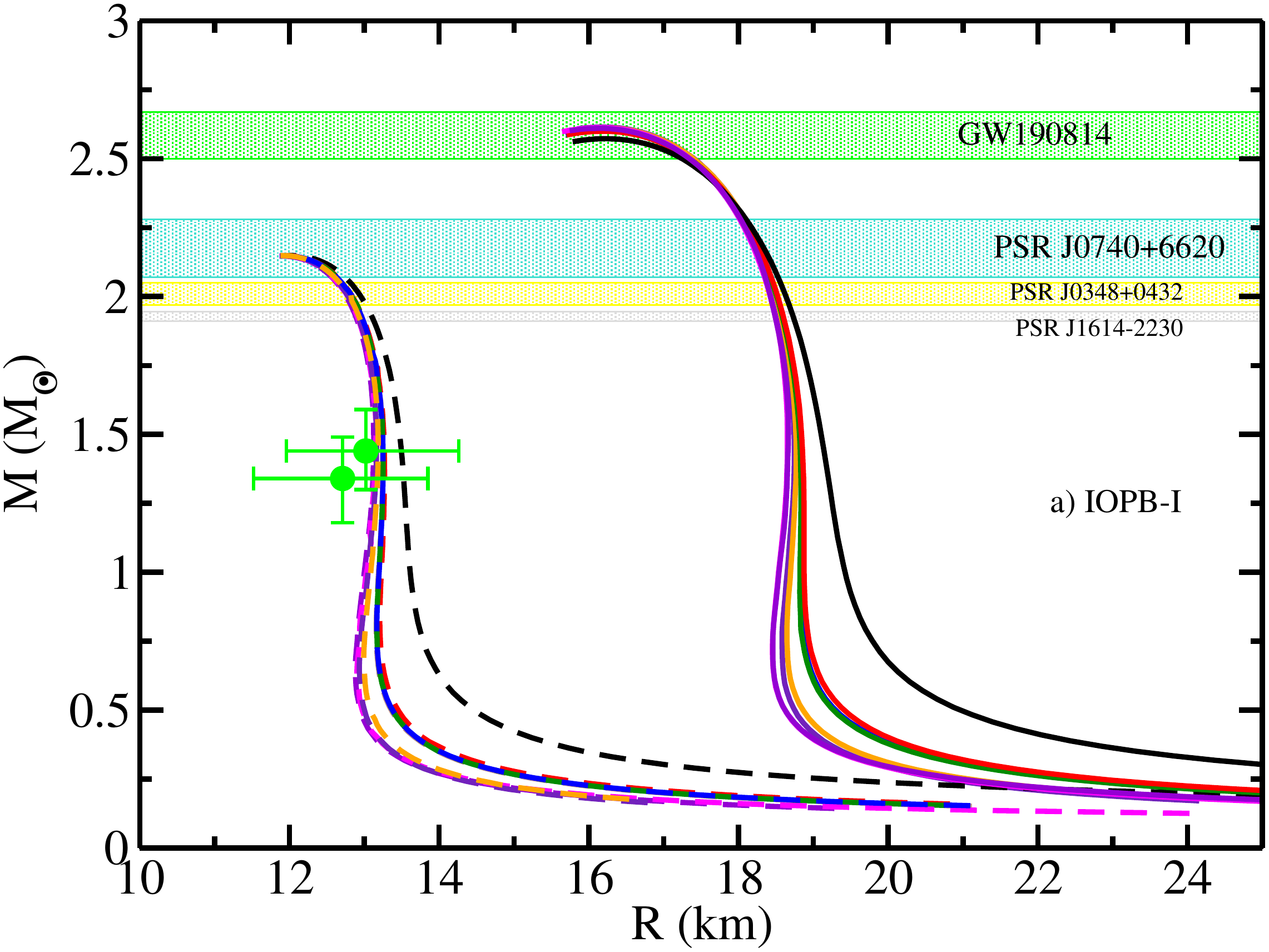}
	\hspace{0.8cm}
	\includegraphics[scale=0.33]{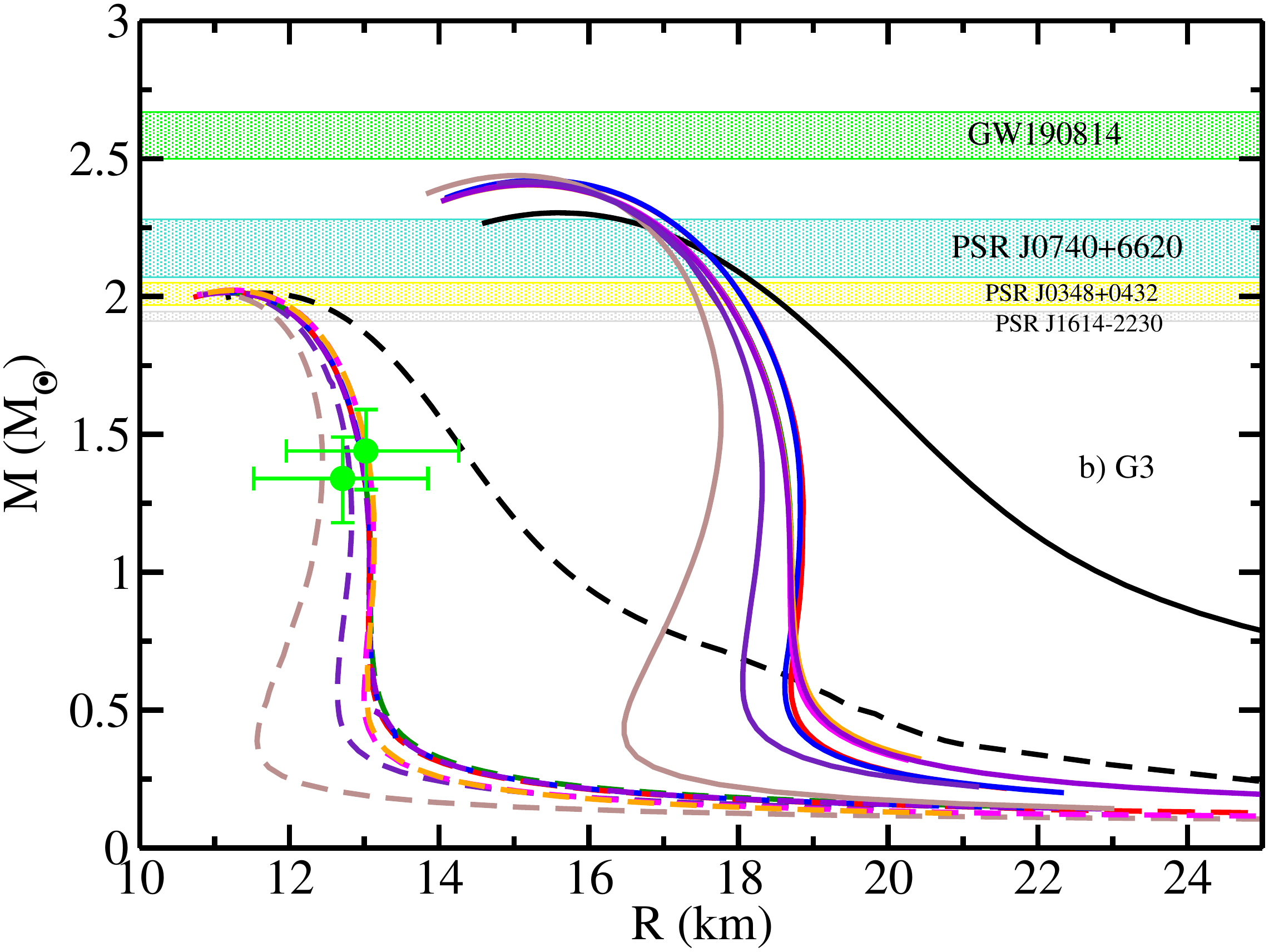}
\end{adjustbox}

\vspace{0.5cm}
From the above MR relations, we see that the maximum mass and the radius of both static and rotating NS's do not change by a large amount using the inner crust with different slope parameters. The unified EoSs (NL3, TM1, and IU-FSU) produce NS's with a small radius at the canonical mass. However, for non-unified EoSs (IOPB-I and G3),  the inner crust EoS from same or different model with smaller symmetry energy slope parameter $L_{sym}$ predicts a smaller radius at canonical mass of a NS.  \par

The effect of inner crust on the static NS tidal deformability $\lambda$ is also studied for all core EoSs. In addition to this, the variation in the RNS properties like Moment of Inertia (MI) is also discussed. The tidal deformability of a NS is defined as \cite{PhysRevD.81.123016,PhysRevC.95.015801}
\begin{equation}\label{l1}
\lambda=-\frac{Q_{ij}}{\mathcal{E}_{ij}} = \frac{2}{3}k_2 R^5,
\end{equation}
where, $Q_{ij}$ is the induced quadrupole mass and $\mathcal{E}_{ij}$ is the external field.
The dimensionless tidal deformability then follows from the $\lambda$ as
\begin{equation}\label{l2}
\Lambda=\frac{\lambda}{M^5}=\frac{2k_2}{3C^5},
\end{equation}
where, $k_2$ is the second love number and $C=M/R$ is the compactness parameter. The expression for the love number is given as \cite{PhysRevD.81.123016}
\begin{eqnarray}\label{l3}
k_2=\frac{8}{5}(1-2C)^2 [2C(y-1)]\Bigl\{2C(4(y+1)C^4
+(6y-4)C^3 
+(26-22y)C^2+3(5y-8)C-3y+6) \nonumber \\
-3(1-2C)^2 
(2C(y-1)-y+2)log\Big(\frac{1}{1-2C}\Big)\Bigr\}^{-1}.
\end{eqnarray}
The value of $y=y(R)$ can be computed by solving the following differential equation \cite{PhysRevC.95.015801,Hinderer_2008}
\begin{equation}\label{l4}
r\frac{dy(r)}{dr}+y(r)^2+y(r)F(r)+r^2 Q(r)=0,
\end{equation}
where,
\begin{equation}\label{l5}
F(r)=\frac{r-4\pi r^3 [\mathcal{E}(r)-P(r)]}{r-2M(r)},
\end{equation}
\begin{eqnarray}\label{l6}
Q(r)=\frac{4\pi r\Big(5\mathcal{E}(r)+9P(r)+\frac{\mathcal{E}(r)+P(r)}{\partial P(r)/\partial\mathcal{E}(r)}-\frac{6}{4\pi r^2}\Big)}{r-2M(r)} 
-4\Bigg[\frac{M(r)+4\pi r^3 P(r)}{r^2 (1-2M(r)/r)}\Bigg]^2.
\end{eqnarray}

\vspace{0.5cm}
\begin{figure}[hbt!]
	\centering
	\includegraphics[height=8cm, width=10cm]{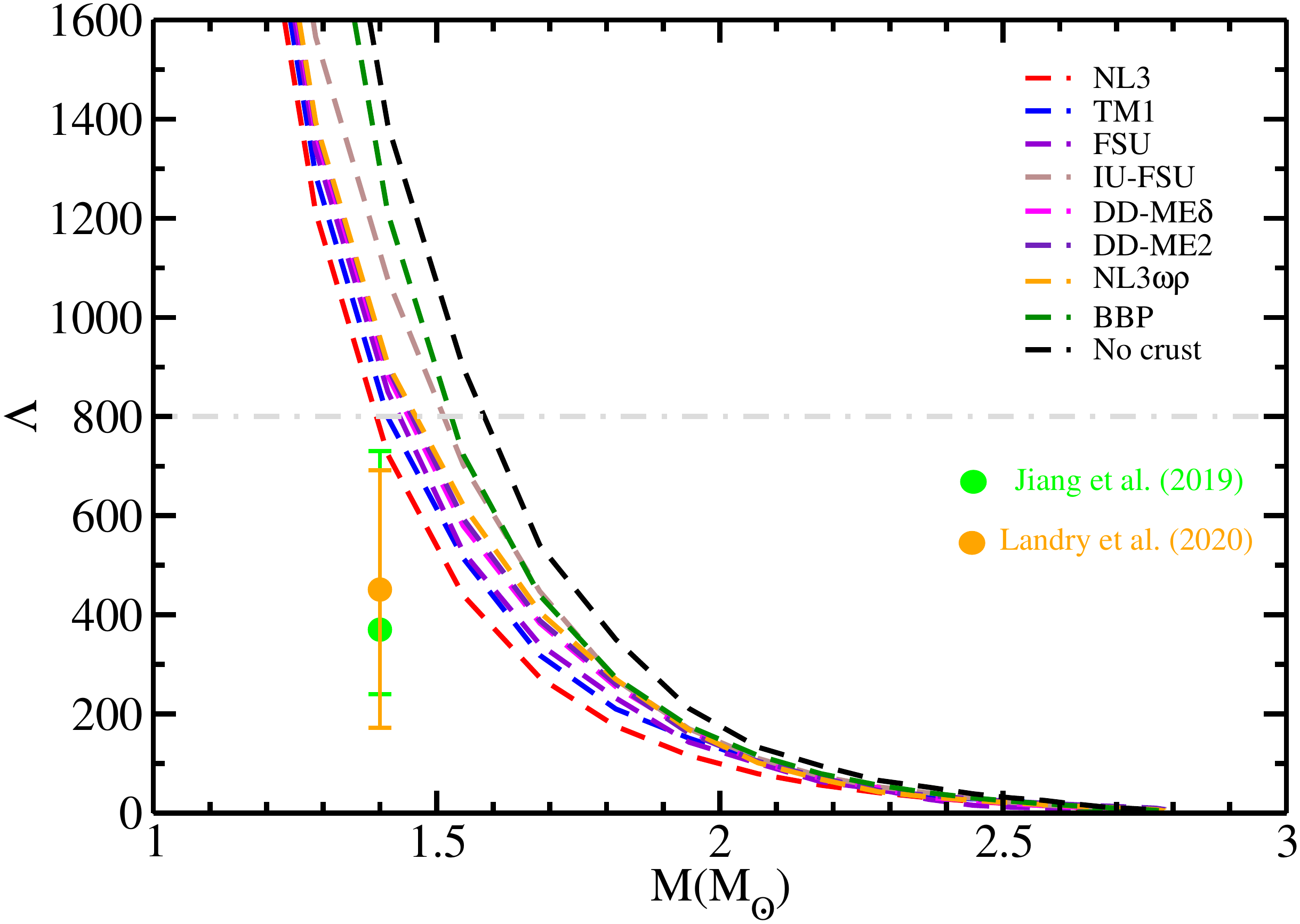}
	\caption{The relation between dimensionless tidal deformability and the mass of a NS for NL3 core EoS with different inner crust EoS. The overlaid arrows (green and orange) represents the combined constraints on the tidal deformability at 1.4$M_{\odot}$ from  PSR J0030+0451 and GW170817 and x ray data \protect\cite{PhysRevLett.121.161101, Jiang_2020,PhysRevD.101.123007} . The grey dashed line represents the upper limit on the $\Lambda_{1.4}$ value \protect\cite{PhysRevLett.119.161101} .}
	\label{FIG:5}
\end{figure}

Figure (\ref{FIG:5}) shows the variation of the dimensionless tidal deformability with the NS mass for NL3 core EoS with different inner crust EoS. The constraints on the $\Lambda$ from the recent GW data is also shown. The green overlaid arrow shows the recent constraints on the $\Lambda_{1.4}$ from the combined data of PSR J0030+0451 and GW170817,  $\Lambda_{1.4}=370_{-130}^{+360}$ \cite{Jiang_2020}, while the orange one shows the non-parametric constraints from PSRs + GWs + x ray, $\Lambda_{1.4}=451_{-279}^{+241}$  \cite{PhysRevD.101.123007}.The grey dashed line represents the upper limit on the dimensionless tidal deformability at the canonical mass from GW170817 data, $\Lambda_{1.4}$=800  \cite{PhysRevLett.119.161101}. The NL3 unified EoS predicts the lowest value of the dimensionless tidal deformability at 1.4$M_{\odot}$, $\Lambda_{1.4}$=800, due to the small radius at the canonical mass. The other non-unified NL3 EoSs (NL3 core + other crust EoSs except NL3) predict a value in the range 800-1400 with $\Lambda_{1.4}$=1400 for NL3 without the inner crust. This shows that the unified EoS is important in determining the NS properties that support the contraints from recent GW data. \par 

\vspace{1.0cm}
\begin{figure}[hbt!]
	\centering
	\includegraphics[width=10cm, height=8cm]{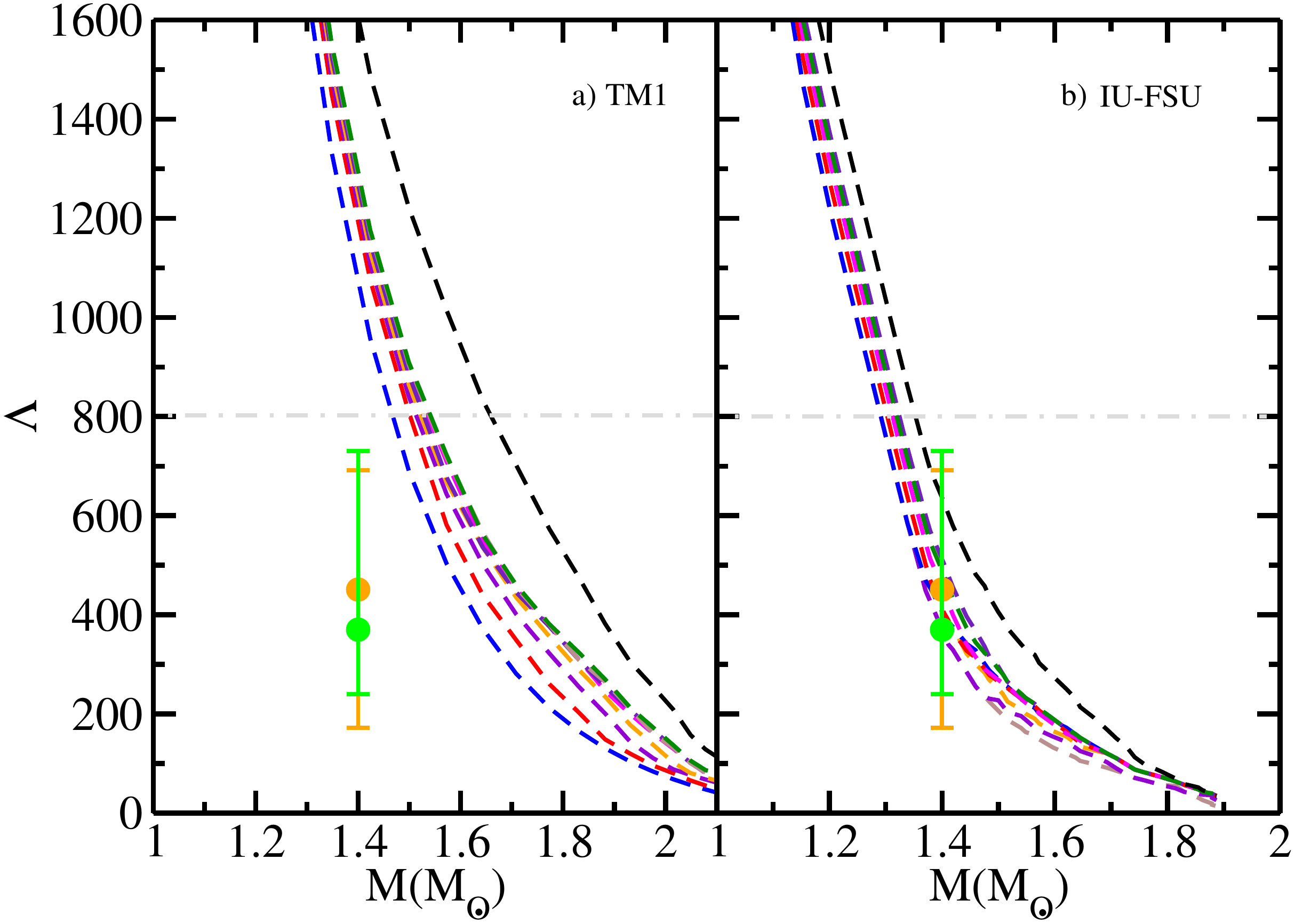}
	\caption{Same as figure (\ref{FIG:5}), but for a) TM1 and b) IU-FSU core EoS.}
	\label{FIG:6}
\end{figure}

\vspace{1.0cm}
\begin{figure}[hbt!]
	\centering
	\includegraphics[width=10cm, height=8cm]{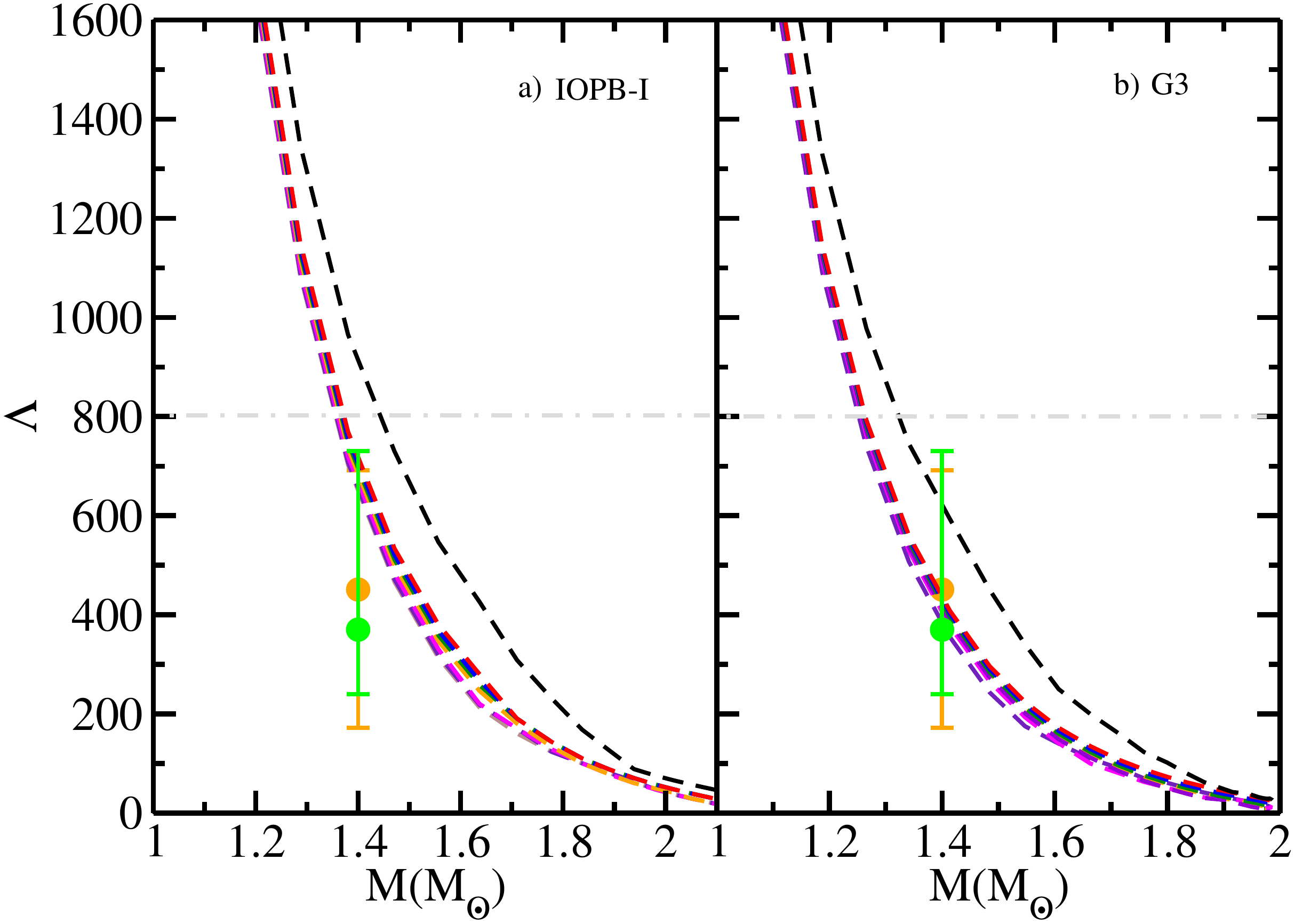}
	\caption{Same as figure (\ref{FIG:5}), but for a) IOPB-I and b) G3 core EoS.}
	\label{FIG:7}
\end{figure}
Figure (\ref{FIG:6}) shows the dimensionless tidal deformability for TM1 and IU-FSU core EoS with different crust EoSs. The unified TM1 EoS (TM1 crust + TM1 core) provides a low value for tidal deformability, $\Lambda_{1.4}$ =1060. This value increases upto to $\Lambda_{1.4}$= 1587 for TM1 NS without inner crust. For IU-FSU, the unified EoS gives $\Lambda_{1.4}$=368, and $\Lambda_{1.4}$=637 without inner crust EoS. For other non-unified EoSs, the variation in the tidal deformability at 1.4$M_{\odot}$ is very small for both TM1 and IU-FSU core EoSs.\\

Figure (\ref{FIG:7}) shows the dimensionless tidal deformability for IOPB-I and G3 core sets. For IOPB-I, the FSU crust EoS predicts the smallest value of $\Lambda_{1.4}$=637, while the other crust EoSs determine the value in the range 640-730. Similarly for  G3 set, the IU-FSU gives a low value for tidal deformability, $\Lambda_{1.4}$=349, while others provide a value in the range 393-450. The value increases with the increase in the value of symmetry energy slope parameter. The NS without inner crust provides a value of $\Lambda_{1.4}$=914 and 620 for IOPB-I and G3 sets respectively.\\

In the simplest form, the moment of inertia $I$ is defined as the ratio of the angular momentum to the angular velocity of a NS, $I=J/\Omega$. In terms of the angular frequency $\omega$, the moment of inertia is defined  as \cite{LATTIMER2000121,Worley_2008}
\begin{equation}
I\approx\frac{8\pi}{3}\int_{0}^{R} (\mathcal{E}+P)e^{-\phi(r)} \Big[1-\frac{2m(r)}{r}\Big]^{-1}\frac{\bar{\omega}}{\Omega}r^4 dr,
\end{equation}
where, $\bar{\omega}$ is the dragging angular velocity of a rotating star, satisfying the boundary conditions
\begin{equation}
\bar{\omega}(r=R)=1-\frac{2I}{R^3}, \frac{d \bar{\omega}}{dr}|_{r=0}=0,
\end{equation}

The slow rotation approximation for the NS moment of inertia neglects the deviation from the spherical symmetry and becomes independent of the angular velocity $\Omega$. The MI for the slowly rotating NS  computed throught the above formalism can be approximated by the following empirical relation \cite{Lattimer_2005}

\begin{equation}
I\approx (0.237 \pm 0.008)MR^2  \times \Bigg[1+4.2 \frac{M}{M_{\odot}} \frac{km}{R} +90 \Bigg(\frac{M}{M_{\odot}} \frac{km}{R}\Bigg)^4\Bigg].
\end{equation}

In the present work, the NSs rotating at the kepler frequency are studied, hence the numerical calculations for the moment of inertia are performed using the RNS code.
The MI of a NS has been calculated by various groups \cite{Stergioulas2003,friedman_stergioulas_2013,Paschalidis2017}, but the variation in the value of MI with different inner crust EoS hasn't been calculated.\par
\vspace{1.0cm}
\begin{figure}[hbt!]
	\centering
	\includegraphics[width=10cm, height=8cm]{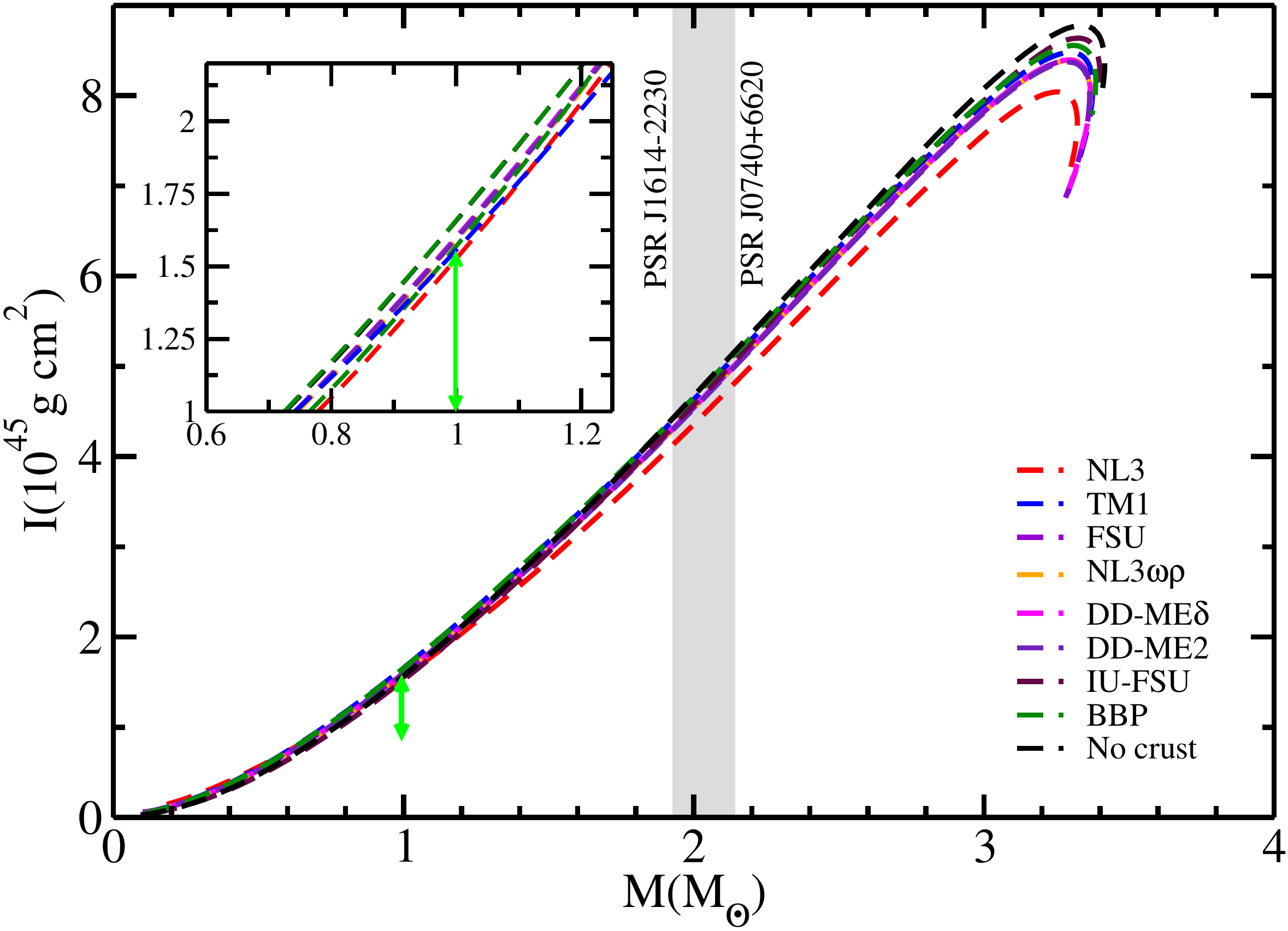}
	\caption{Variation of moment of inertia $I$ with the mass of a NS for NL3 core EoS with different crust EoS. The green arrow represents the constraints on the moment of inertia from PSR J0737-3039A \protect\cite{Landry_2018,PhysRevD.99.123026}obtained from the GW 170817 data analysis \protect\cite{PhysRevLett.119.161101,PhysRevLett.121.161101} . }
	\label{FIG:8}
\end{figure}

The variation in the moment of inertia for RNS with NL3 core EoS and different crust EoSs is shown in figure (\ref{FIG:8}). As clear from the figure, the change in the moment of inertia with different crust EoSs is very small. For unified NL3 EoS, the value of $I$ is found to be 1.53$\times$ 10$^{45}$g cm$^2$ well satisfying the constraint from PSR J0737-3039A  $I$=1.53$_{-0.24}^{+0.38}\times$ 10$^{45}$g cm$^2$. For the non-unified EoSs, the moment of inertia increases with the symmetry energy slope parameter $L_{sym}$ as they predict a large radius.\par
\begin{figure}[hbt!]
	\centering
	\includegraphics[width=10cm, height=8cm]{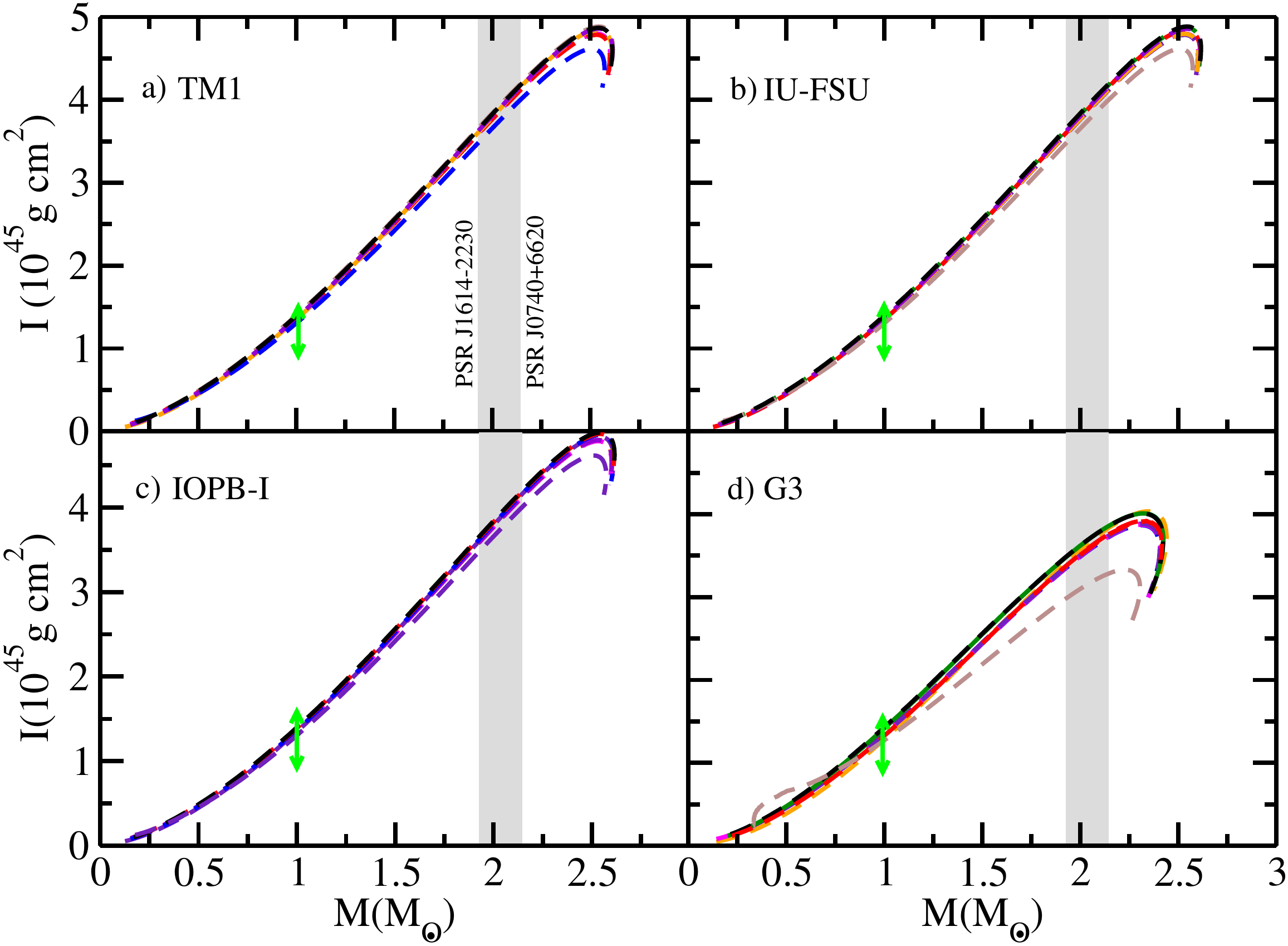}
	\caption{Same as figure (\ref{FIG:8}), but for a) TM1, b) IU-FSU, c) IOPB-I , and d) G3 core EoSs. }
	\label{FIG:9}
\end{figure}

For TM1, IU-FSU, IOPB-I, and G3 core EoSs, the moment of inertia variation with the NS mass is shown in figure (\ref{FIG:9}). For the TM1 and IU-FSU, the unified EoS provides a small moment of inertia $I$=1.31 \& 1.29$\times$ 10$^{45}$g cm$^2$, respectively. With no unified EoS available for IOPB-I and G3 sets, the low symmetry energy slope parameter crust EoS provides a lower value of the moment of inertia, $I$=1.27 \& 1.22$\times$ 10$^{45}$g cm$^2$, respectively. The moment of inertia is approximated by the relation $I\propto MR^2$ which shows that it increases almost linearly with the NS mass for all models. The NS radius starts to decrease as soon as the maximum mass is achieved which allows the moment of inertia to drop sharply. Since $I$ is proportional to the mass linearly and square of the radius, it is more sensitive to the density dependence of nuclear symmetry energy and its derivatives like slope parameter, which influence the NS radius \cite{Worley_2008,Li2019}. By using different inner crust equation of states for a given core EoS, the change in the radius at the canonical mass is observed, which affects the moment of inertia. Therefore the contribution to the moment of inertia due to the crust part of the star is much less than the core part.\par
\vspace{1.0cm}

To study the effect of different crust EoSs on the rns properties, we have calculated the kepler frequency for NL3, TM1, and IU-FSU parameter sets only as they contain both unified as well as non-unified EoSs. \par
\vspace{1.0cm}
\begin{figure}[hbt!]
	\centering
	\includegraphics[width=10cm, height=8cm]{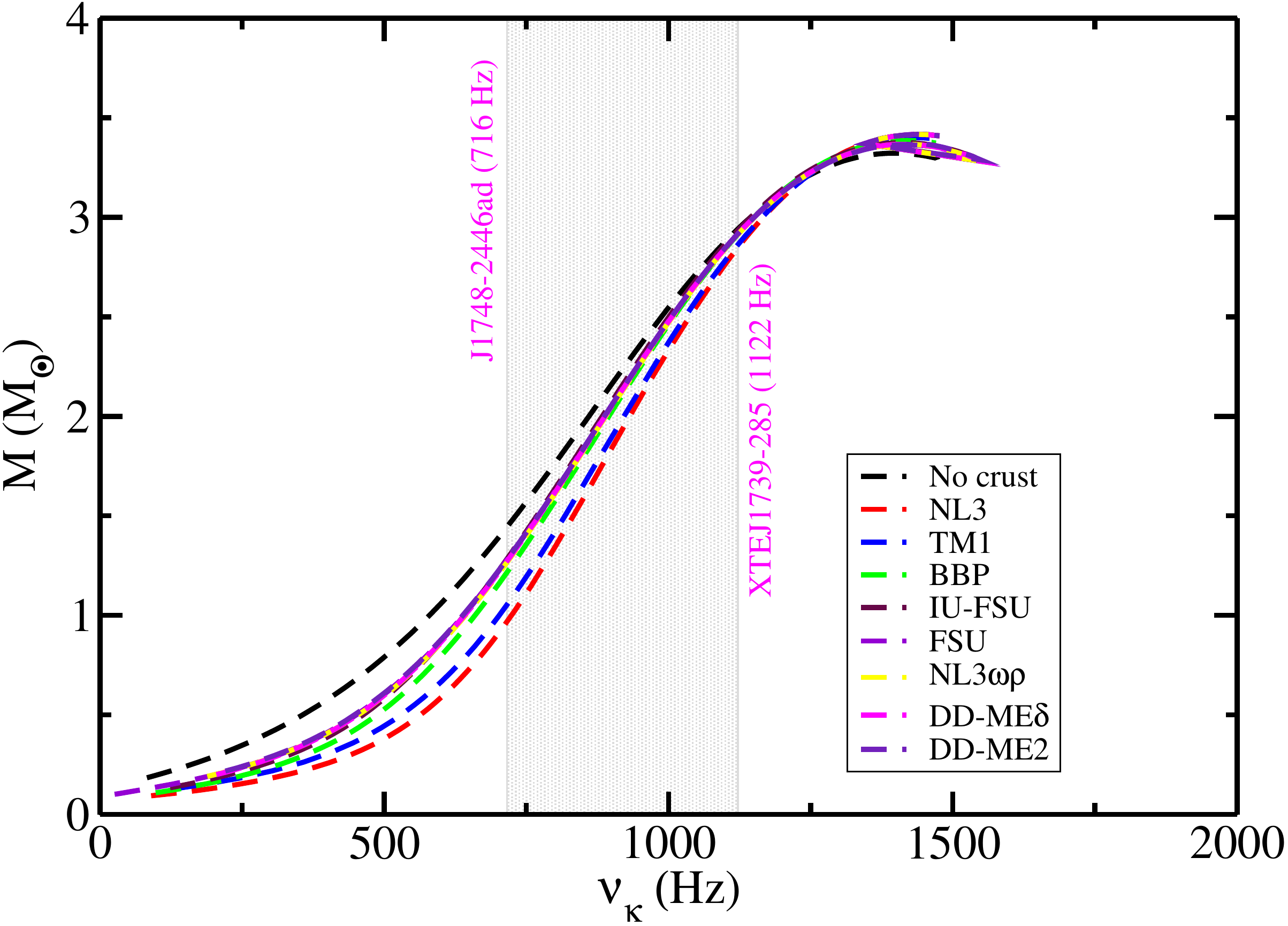}
	\caption{Kepler frequency $\nu_k$ vs the gravitational mass of NS for NL3 parameter set using different crust EoSs. The grey band represents the observational limits on the frequency from rapidly rotating pulsars PSR J1748-2446ad ($\nu$=716 Hz) and XTE J1739-285 ($\nu$=1122 Hz). }
	\label{FIG:10}
\end{figure}  
Figure (\ref{FIG:10}) shows the NS gravitational mass as a function of kepler frequency. Since the maximum mass obtained for the NL3 core with different crust EoSs does not vary too much, the frequency at the maximum mass remains almost same for all equation of states. The maximum frequency lies around 1400 Hz. With different crust EoS, the variation at the 1.4$M_{\odot}$ is observed. For the EoS with NL3$_{crust}$ +NL3$_{core}$, the frequency at the canonical mass is found to be $\approx$ 800 Hz. For other crust EoSs, the frequency at 1.4$M_{\odot}$ is less than 800 Hz with around 700 Hz for EoS without inner crust. This variation in the frequency with different inner crust follows from the change in the radius of NS, which is affected by the change in the symmetry energy slope parameter of the crust EoS.  \par
\vspace{1.0cm} 
\begin{figure}[hbt!]
	\centering
	\includegraphics[width=10cm, height=8cm]{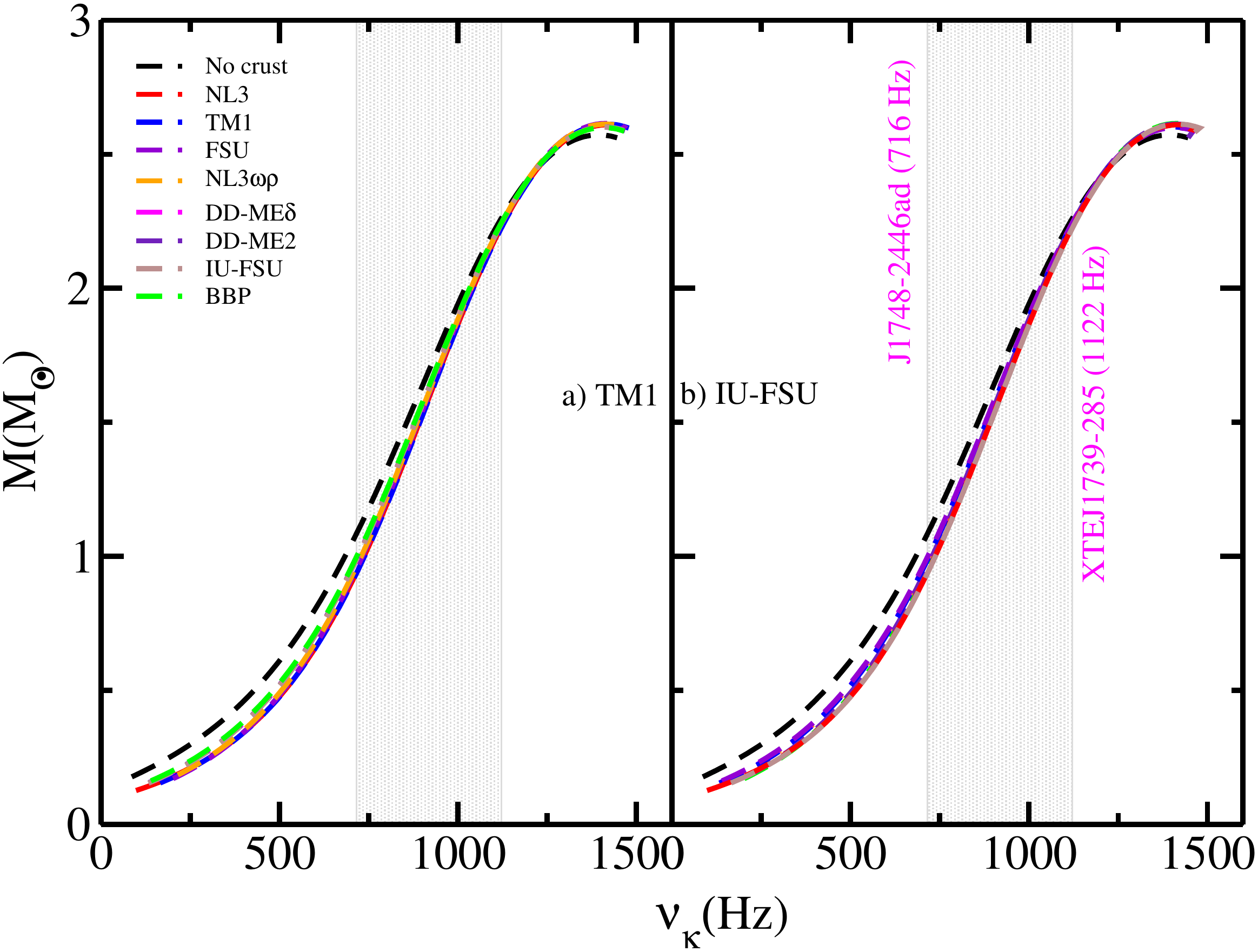}
	\caption{Same as fig. (\ref{FIG:10}), but for a) TM1 and b) IU-FSU parameter sets. }
	\label{FIG:11}
\end{figure}
Figure (\ref{FIG:11}) shows the kepler frequency variation with the NS mass using different crust EoS for TM1 and IU-FSU parameter sets. For TM1 and IU-FSU, all the EoSs obtained rotate with a maximum frequency of $\approx$ 1400 Hz. The variations at the canonical mass are also too small. The EoS obtained without the inner crust part rotates with a frequency of around 820 Hz at 1.4$M_{\odot}$ for both TM1 and IU-FSU sets. Thus we see that the change in the symmetry energy slope parameter of crust EoS doesn't affect the rotational frequency of the rns directly. \par
\begin{figure}[hbt!]
	\centering
	\includegraphics[width=10cm, height=8cm]{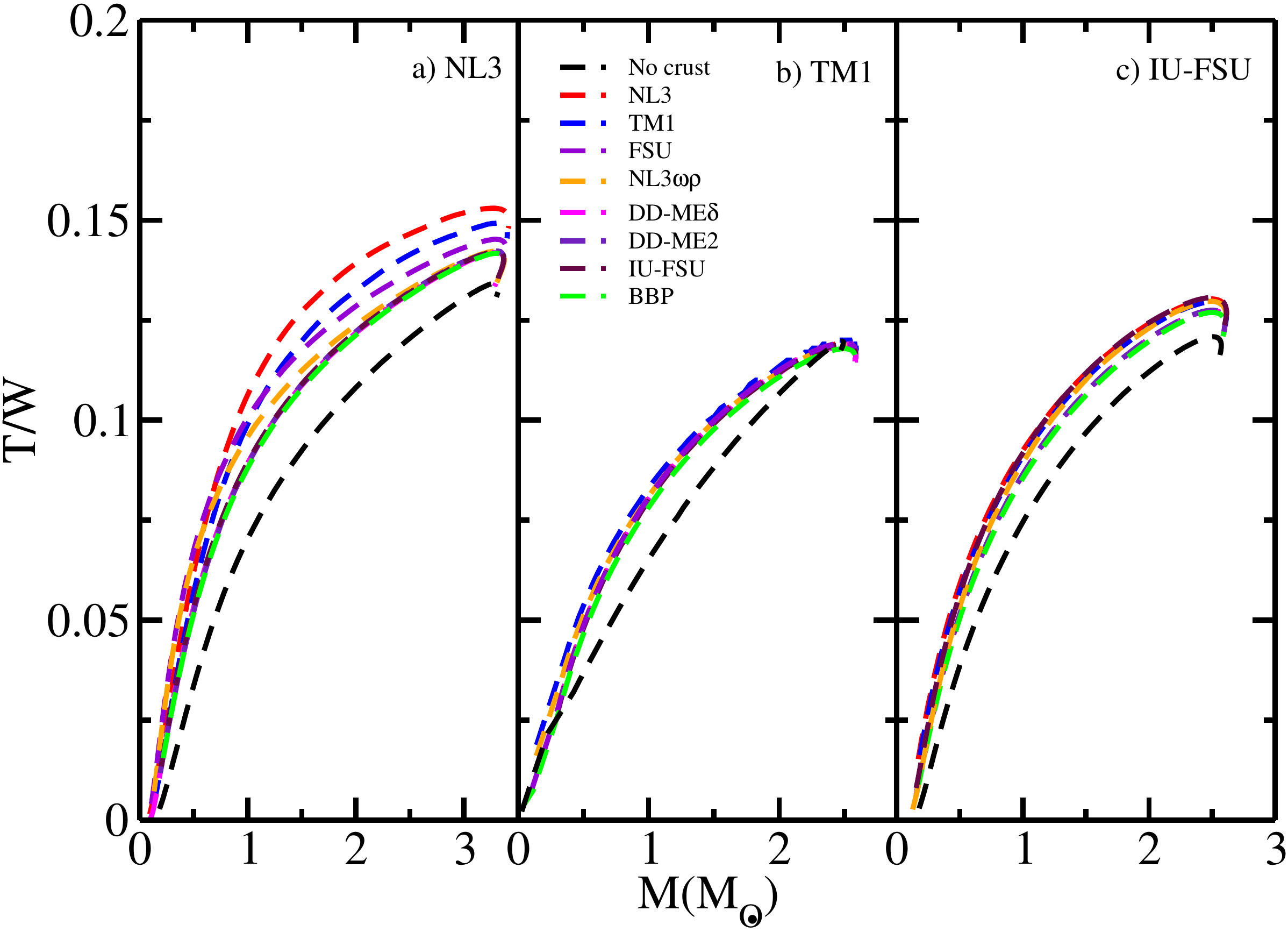}
	\caption{The variation of the Rotational kinetic energy to the gravitational potential energy ratio with the NS mass for a) NL3, b) TM1 and c) IU-FSU parameter sets with different crust EoSs. }
	\label{FIG:12}
\end{figure}

   
An important quantity that characterizes the rotation of a star is the T/W ratio, which is defined as the ratio of the rotational kinetic energy T to the gravitational potential energy W. For a given rotating star, if this ratio is greater than a critical value, the star becomes dynamically unstable. The critical value of T/W ratio is not fixed. Some measurements predict a value of 0.27 as the critical limit \cite{10.1093/gji/21.1.103-a,1985ApJ...298..220T} and some show it to be in the range 0.14-0.27 \cite{1996ApJ...458..714P,1995ApJ...444..363I,2001ApJ...550L.193C}. Figure (\ref{FIG:12}) displays the  variation in T/W ratio of a rns with the gravitational mass. With the increase in the central density, the angular velocity increases which in turn produces a star with higher value of T/W ratio. The unified EoS for NL3, TM1, and IU-FSU parameter sets predict the highest value of the T/W ratio as compared to other non-unified equation of states. For NL3 set, the unified EoS (L${crust}$=L$_{core}$= 118.3 MeV) has T/W ratio of 0.15, while for TM1 and IU-FSU unified EoSs, the ratio is 0.12 and 0.13, respectively. The EoS without inner crust part predicts a value of 0.13, 0.11, and 0.12 for NL3, TM1, and IU-FSU sets respectively.\par 
\vspace{1.0cm}

From the measurements of the rns properties like the moment of inertia, frequency, and T/W ratio for NL3, TM1, and IU-FSU unified as well as non-unified equation of states, we see that such quantities vary with a change in the symmetry energy slope parameter $L_{sym}$ of the inner crust part, but such variation is more proportional to the NS mass and radius. A change in the NS radius at the canonical mass alters the structure of the star which influences the other properties of the star. A relation between the slope parameter and the above defined rns properties for the inner crust part will allow us to study these properties in great detail leading to a proper choice of the crust EoS.\par 

\begin{table}[ht]
	\centering
	\caption{Variation (in percent) in the maximum mass, corresponding radius, and the radius at 1.4$M_{\odot}$ of a NS without inner crust and the corresponding inner crust.} 
	{\begin{tabular}{p{1.0cm}p{1.2cm}cccccccc} 
			Model &  & BBP & IU-FSU & DD-ME2 & DD-ME$\delta$ & NL3$\omega \rho$ & FSU & TM1 & NL3 \\
			\hline			
			
			&  $\Delta M_{max}$ & 0.25&0.14&0.11&0.07&0.11&0.11&0.32&0.58\\
			NL3 & $\Delta R_{max}$ & 1.90&2.62&0.87&1.22&0.93&0.79&0.73&1.19\\
			& $\Delta R_{1.4}$ & 3.28&3.12&3.22&3.30&3.54&4.74&6.51&7.57\\
			\hline
			
			&  $\Delta M_{max}$ & 0.60&1.65&1.51&1.70&1.38&1.56&0.64&1.65\\
			TM1 & $\Delta R_{max}$ & 4.51&2.12&1.85&1.92&4.68&2.98&4.46&3.29\\
			& $\Delta R_{1.4}$ & 9.69&10.22&10.19&10.16&10.13&10.11&12.73&11.28\\
			\hline
			
			&  $\Delta M_{max}$ & 0.10&0.00&0.05&0.26&0.10&0.05&0.15&0.21\\
			IU-FSU & $\Delta R_{max}$ & 1.35&2.07&1.41&2.23&1.25&1.26&1.71&1.79\\
			& $\Delta R_{1.4}$ & 2.91&3.78&3.55&3.42&3.30&3.19&2.78&2.53\\
			\hline
			
			&  $\Delta M_{max}$ &0.04&0.02&0.06&0.04&0.04&0.07&0.04&0.04\\
			IOPB-I & $\Delta R_{max}$ &1.11&1.17&0.66&0.61&0.73&0.54&0.92&1.30 \\
			& $\Delta R_{1.4}$ & 1.84&2.57&2.34&2.29&2.26&2.89&1.85&1.79\\
			\hline
			
			&  $\Delta M_{max}$ & 0.05&0.00&0.15&0.20&0.15&0.30&0.04&0.50\\
			G3 & $\Delta R_{max}$ & 3.94&5.74&4.02&3.79&4.24&3.28&0.92&3.50\\
			& $\Delta R_{1.4}$ & 11.50&13.91&11.82&11.66&9.45&11.36&9.56&10.17\\
			\hline
			\hline
		\end{tabular}\label{tbl2}}
\end{table}

Table \ref{tbl2} shows the deviation in the properties of a static NS like maximum mass, the corresponding radius, and the radius at the canonical mass for the given parameter sets. The deviations are calculated by considering the NS with different inner crust EoS with respect to the NS without inner crust. The NS without inner crust predicts a very large radius at the canonical mass, while the unified EoS (same crust and core EoS) for any parameter set gives the low radius NS, which satisfies the radius constraints as explained before. The deviation between a NS without inner crust and with crust is large, which implies that the unified EoS gives a better estimate of the radius at the canonical mass as compared to other non-unified EoSs. It is clear that the variation in the maximum mass and the corresponding radius are very small for EoSs, but the radius at 1.4 $M_{\odot}$ is highly impacted by the inner crust EoS. For NL3 core EoS, the variation in the radius at $R_{1.4}$ is maximum for NL3 inner crust $\approx$ 7.57$\%$ which is around 1.2 km. The maximum variation in the radius, $R_{1.4}$, for IOPB-I and G3 core EoSs is with the FSU and IU-FSU inner crust EoS, respectively. Such large deviations in the radius, $R_{1.4}$ show that a proper choice of inner crust EoS is important to calculate the mass and radius of a NS with small uncertainities in these values. The unified EoSs for NL3, TM1, and IU-FSU predict small values of radius, tidal deformability, at the canonical mass. For RNS, the radius at 1.4$M_{\odot}$ varies with different crust EoS. The moment of inertia doesn't vary largely. For the core EoSs without having a unified EoS, the crust with a smaller symmetry energy slope parameter $L_{sym}$ predicts small values of radius, the tidal deformability for SNS and radius, moment of inertia for RNS. Thus the crust-core transition allows the construction of a stellar EoS and a precise measurement of the NS properties for both static and rotating stars.\\

The constraints on the inner crust EoS of a NS and the proper matching of inner crust with core EoS is helpful in considering the nuclear and astrophysical applications of the RMF model. A core EoS with a smaller symmetry energy slope parameter implies small symmetry energy at high densities \cite{Zhang2019}. For a model with higher symmetry energy at sub-saturation density, the inner crust properties of a NS are affected in addition to the pasta phases as shown in refs. \cite{PhysRevC.90.045802,PhysRevC.89.045807}. Studies have shown that for a complete unified EoS, the inner crust part should either be from the same model or the symmetry energy slope parameter should match. However, as we have seen in the above plots, the inner crust EoS with smaller slope parameter $L_{sym}$ predicts a smaller radius at 1.4 $M_{\odot}$ as compared to the one with large $L_{sym}$. The same trend is followed by all parameter sets with unified EoS. For non-unified EoS, the crust from same or different model but with smaller symmetry energy slope parameter gives a low radius NS. While the inner crust does affect the radius of RNS, the moment of inertia varies only by a small factor.\par 

\section{Summay and conclusion}
\label{sec:3}
The NS properties like mass and radius were investigated using the relativistic mean-field (RMF) model. To study the effect of symmetry energy and its slope parameter on a NS, the inner crust EoSs with different symmetry energy slope parameters have been used. For the outer crust, the BPS EoS is used for all sets as the outer crust part doesn't affect the NS maximum mass and radius. For the inner crust part, the NL3, TM1, FSU, NL3$\omega \rho$, DD-ME$\delta$, DD-ME2, and IU-FSU parameter sets have been used whose slope parameter varies from 118.3-47.2 MeV. For the core part, NL3, TM1, IU-FSU, IOPB-I, and G3 parameter sets are used. The unified EoSs are constructed by properly matching the inner crust EoS with outer crust and core EoS using the thermodynamic method. The EoSs constructed for the spherical and symmetrical NS under charge neutral and $\beta$-equilibrium conditions are taken as the input into the TOV equation to obtain NS properties. It is seen that although the NS maximum mass and the corresponding radius do not change by a large amount, the radius at the canonical mass, $R_{1.4}$ are largely impacted by using inner crust EoSs with different symmetry energy slope parameter. By varying the slope parameter from low to high values, the radius $R_{1.4}$ also increases. Parameter sets with different nuclear matter properties for the NS core. The effect of $L_{sym}$ on the NS maximum mass, radius, and the radius at 1.4$M_{\odot}$ are calculated and the variation of about 2 km is found in the radius at the canonical mass. The properties like mass, radius, the moment of inertia, T/W ratio of a maximally rotating stars are also calculated using same combination of EoSs. It is seen that similar to SNS, the maximum mass and the corresponding radius for RNS do not vary much, but the radius at the canonical mass is affected by the slope parameter. The moment of inertia doesn't vary too much with change in the symmetry energy slope parameter $L_{sym}$. The kinetic to potential energy ratio also varies with the change in the symmetry energy slope parameter of the crust. Such rns properties are more related to the mass and the radius of the star and the variation in the radius at the canonical mass influences the other properties of the star. \\
There are several different aspects that need to be further studied in the current work. A unified EoS for the parameter sets like IOPB-I and G3 with both crust and core part described by the same model with different slope parameter $L_{sym}$ will be a better investigation to see the behavior of radius and other NS properties at canonical mass. 

	\section*{Acknowledgement}
	We are thankful to the referee for enhancing our understanding on the subject and the quality of the paper.	
	A.A.U. acknowledges the Inter-University Centre for Astronomy and Astrophysics, Pune, India for support via an associateship and for hospitality.

\bibliography{mybibfile}

\end{document}